\pdfoutput=1  

\documentclass[aps,prd,superscriptaddress,floatfix,nofootinbib,preprintnumbers,eqsecnum,twocolumn]{revtex4-1}

\usepackage[utf8]{inputenc}
\usepackage[dvips]{graphicx}
\usepackage[dvipsnames]{xcolor}

\usepackage{amsmath,float,graphicx,placeins,amssymb,multirow,pgfplots,pgfplotstable}
\usepackage{empheq}
\usepackage{tikz}
\usepackage{comment}
\usepackage{enumerate,slashed,cancel,comment,bbm,rotating,placeins,mathtools,multirow,hhline}
\usepackage[normalem]{ulem}
\usepackage{array}

\usepackage{epstopdf}
\usepackage{hyperref}
\usepackage{mathrsfs}
\usepackage{amssymb}
\usepackage{physics}
\usepackage{booktabs}

\usepackage{color}
\usepackage{relsize}
\usepackage{graphics}
\usepackage{epstopdf}
\usepackage{hyperref}
\usepackage{mathrsfs}
\usepackage{amssymb}
\usepackage{physics}
\usepackage{booktabs}

\usepackage[normalem]{ ulem }
\usepackage{amsthm}
\usepackage{amsmath}
\usepackage{cancel}
\usepackage{float}
\usepackage{fontawesome} 
\usepackage[caption=false]{subfig}
\usepackage{tabularx,ragged2e} 
 \newcolumntype{L}{>{\RaggedRight\arraybackslash}X}
 
\colorlet{mylinkcolor}{ForestGreen}
\colorlet{mycitecolor}{Red}
\colorlet{myurlcolor}{violet}

\usepackage{lipsum}
\usepackage{diagbox}
\usetikzlibrary{shapes.geometric,arrows}
\graphicspath{}

\def\beq{\begin{equation}}
\def\eeq{\end{equation}}
\def\beqn{\begin{eqnarray}}
\def\eeqn{\end{eqnarray}}

\hypersetup{
 linkcolor = {blue!80!black},
 citecolor = {blue!70!black},
 urlcolor = {blue!80!black},
 colorlinks = true
}

\def\ie{{\it i.e.}\/}
\def\eg{{\it e.g.}\/}
\def\cf{{\it cf.}\/}
\def\etc{{\it etc}.\/}

\def\IR{\relax{\rm I\kern-.18em R}}
 \font\cmss=cmss10 \font\cmsss=cmss10 at 7pt
\def\IQ{\relax{\rm I\kern-.18em Q}}
\def\IZ{\relax\ifmmode\mathchoice
 {\hbox{\cmss Z\kern-.4em Z}}{\hbox{\cmss Z\kern-.4em Z}}
 {\lower.9pt\hbox{\cmsss Z\kern-.4em Z}}
 {\lower1.2pt\hbox{\cmsss Z\kern-.4em Z}}\else{\cmss Z\kern-.4em Z}\fi}

\newcommand{\vecbf}[1]{\vec{\mathbf{#1}}}

\pgfplotsset{compat=1.17}
\begin{document}


\title{Conversations and Deliberations: \\ Non-Standard Cosmological Epochs
and Expansion Histories 
}

\def\andname{\hspace*{-0.5em}} 

\author{Brian Batell}
\email[Email address: ]{batell@pitt.edu}
\affiliation{Pittsburgh Particle Physics, Astrophysics, and Cosmology Center, 
Department of Physics and Astronomy, University of Pittsburgh, Pittsburgh, PA 15217, USA}

\author{Keith R.~Dienes}
\email[Email address: ]{dienes@arizona.edu}
\affiliation{Department of Physics, University of Arizona, Tucson, AZ 85721, USA}
\affiliation{Department of Physics, University of Maryland, College Park, MD 20742, USA}

\author{Brooks Thomas}
\email[Email address: ]{thomasbd@lafayette.edu}
\affiliation{Department of Physics, Lafayette College, Easton, PA  18042, USA}

\author{Scott Watson}
\email[Email address: ]{gswatson@syr.edu}
\affiliation{Department of Physics, Syracuse University, Syracuse, NY 13244, USA}
\affiliation{Department of Physics and Astronomy, University of South Carolina, Columbia, SC 29208, USA}

\collaboration{Editors}

\author{Rouzbeh Allahverdi}
\affiliation{Department of Physics and Astronomy, University of New Mexico, Albuquerque, NM 87106, USA}

\author{Mustafa Amin}
\affiliation{Department of Physics and Astronomy, Rice University, Houston, TX 77005, USA}

\author{Kimberly K.~Boddy}
\affiliation{Texas Center for Cosmology and Astroparticle Physics, 
Weinberg Institute for Theoretical Physics, Department of Physics, 
The University of Texas at Austin, Austin, TX 78712, USA}

\author{M.~Sten Delos}
\affiliation{Carnegie Observatories, 813 Santa Barbara Street, Pasadena, CA 91101, USA}

\author{Adrienne L.~Erickcek}
\affiliation{Department of Physics and Astronomy, University of North Carolina at Chapel Hill,
Phillips Hall CB3255, Chapel Hill, NC 27599, USA}

\author{\\ Akshay Ghalsasi}
\affiliation{Jefferson Physical Laboratory, Harvard University, Cambridge, MA 02138, USA}

\author{John T.~Giblin, Jr.}
\affiliation{Department of Physics, Kenyon College, Gambier, OH 43022, USA}
\affiliation{Department of Physics, Case Western Reserve University, Cleveland, OH 44106, USA}
\affiliation{Center for Cosmology and AstroParticle Physics (CCAPP) and Department of Physics, Ohio State University, Columbus, OH 43210, USA}

\author{James Halverson}
\affiliation{Department of Physics, Northeastern University, Boston, MA 02115, USA}
\affiliation{NSF AI Institute for Artificial Intelligence and Fundamental Interactions}

\author{Fei Huang}
\affiliation{Department of Particle Physics and Astrophysics, 
Weizmann Institute of Science, Rehovot 7610001, Israel}

\author{Andrew J.~Long}
\affiliation{Department of Physics and Astronomy, Rice University, Houston, TX 77005, USA}

\author{\\ Lauren Pearce}
\affiliation{Department of Physics and Astronomy, James Madison University, 
Harrisonburg, VA 22807, USA}

\author{Barmak Shams Es Haghi}
\affiliation{Texas Center for Cosmology and Astroparticle Physics, 
Weinberg Institute for Theoretical Physics, Department of Physics, 
The University of Texas at Austin, Austin, TX 78712, USA}

\author{Jessie Shelton}
\affiliation{Illinois Center for Advanced Studies of the Universe and Department of Physics,
University of Illinois at Urbana-Champaign, Urbana, IL 61801, USA}

\author{Gary Shiu}
\affiliation{Department of Physics, University of Wisconsin-Madison
1150 University Avenue, Madison, WI 53706, USA}

\author{Kuver Sinha}
\affiliation{Homer L.\ Dodge Department of Physics and Astronomy, 
University of Oklahoma, Norman, OK 73019, USA}

\author{Tristan L.~Smith}
\affiliation{Department of Physics and Astronomy, Swarthmore College, Swarthmore,
PA 19081, USA}

\collaboration{Discussion Leaders} 

\begin{abstract}
This document summarizes the discussions which took place during the PITT-PACC Workshop 
entitled ``Non-Standard Cosmological Epochs and Expansion Histories,'' held in 
Pittsburgh, Pennsylvania, Sept.~5--7, 2024.  Much like the non-standard 
cosmological epochs that were the 
subject of these discussions, the format of this workshop was also non-standard.
Rather than consisting of a series of talks from participants, with each person presenting 
their own work, this workshop was instead organized around free-form discussion blocks, with each
centered on a different overall theme and guided by a different set of Discussion Leaders.
This document is not intended to serve as a comprehensive review of these topics, but rather 
as an informal record of the discussions that took place during the workshop, in the hope 
that the content and free-flowing spirit of these discussions
may inspire new ideas and research directions. 
\end{abstract}
\maketitle

\tableofcontents


\section*{Preface\label{sec:intro}}


While much is known about the expansion history of the universe since the beginning of 
the Big-Bang nucleosynthesis (BBN) epoch, far less is known about the expansion history at 
earlier times.  As a result, there are many competing possibilities for what might have
happened prior to BBN.~
Motivated by these considerations, the authors of this document convened in Pittsburgh, 
Pennsylvania, Sept.\ 5--7, 2024, in order to take part in a workshop on non-standard 
cosmological epochs and expansion histories hosted by the Pittsburgh Particle Physics Astrophysics 
and Cosmology Center (PITT-PACC) at the University of Pittsburgh.  The goal of this workshop was 
to bring together people working on different aspects of these broad themes in the hope that 
the resulting collective interactions and discussions might not only  spark new ideas concerning
how to probe the first few seconds of the history of the universe, but also help broaden 
our perspectives on how modifications of the cosmological expansion history might manifest 
themselves observationally as well as establish connections or correlations between different
observational probes of early-universe cosmology.  

The format of this workshop was somewhat non-traditional.  Rather than consisting of a series of talks 
delivered by participants on their own work, this workshop was instead organized as a set of 
thematic {\it discussion blocks}\/, with each focused on a particular topic related to the 
overall themes of the workshop.  
Some of these blocks were focused on possible observational signals of modified expansion 
histories (\eg, gravitational-wave signals), while others were focused on particular non-standard 
cosmological epochs (\eg, EMDEs or stasis epochs).  Still others were focused around top-down 
motivations for modifications of the expansion history (\eg, from formal theory). 
At the beginning of each block, one or two designated Discussion Leaders set the stage 
for the ensuing conversation by delivering a short presentation which provided an overview of the 
topic and identified open physics questions and discussion points related to that topic.  
These Discussion Leaders then moderated the free-flowing discussion that followed, for which 
a full hour or more was allotted for each block.  The emphasis was not on these presentations, 
but on the ensuing conversations.

This document does not aim to be a comprehensive review of non-standard cosmological epochs
and expansion histories.  Indeed, there already exist reviews along these 
lines~\cite{Allahverdi:2020bys,Abdalla:2022yfr}.  Rather, this document is intended to capture 
the thrust of the informal conversations which took place during the workshop, in the hope that a 
record of these conversations might help to spark further ideas related to these topics.
In particular, Sect.~\ref{sec:block1} through Sect.~\ref{sec:block9} of this document 
contain records of the discussions which took place during the nine thematic blocks into 
which the workshop was divided.  Sect.~\ref{sec:summary} then contains a record of the 
discussion which took place during a concluding session at the end of the workshop.


\section*{Prologue: ~Setting the Stage\label{sec:background}}


We know a lot about the history of the universe at late times.  Indeed,
measurements of the cosmic microwave background (CMB)~\cite{Planck:2018vyg}, 
the abundances of light elements, the spatial distribution of matter within the 
observable universe, and a variety of other astrophysical observables have provided us with a 
consistent picture of how our universe has evolved since the beginning of the BBN epoch.

By contrast, we know a lot less about the history of the universe at times before the 
BBN epoch.  Nevertheless, some aspects of that early history are crucial for addressing 
unanswered questions in cosmology.  For example, observations indicate that our universe 
has an energy density extremely close to the critical density and exhibits a far greater 
degree of uniformity on large distance scales than one might na\"{i}vely have predicted 
based on the assumption that the universe was radiation-dominated (RD) at all times prior 
to BBN.  Theoretical proposals (\eg, cosmic inflation) involving additional dynamics 
at times before BBN have been advanced in order both to explain the flatness and uniformity 
of the observable universe and to account for the small deviations from perfect homogeneity 
that we observe (\eg, in the CMB).  We do not yet know which --- if any --- of these proposals 
is realized in nature, or precisely how or when the dynamics associated with that proposal 
unfolded.  

One possibility is that the universe underwent a period of accelerated expansion 
(corresponding to an equation-of-state parameter in the range $w < -1/3$) at very early 
times, after which the universe subsequently entered a period of radiation domination 
($w = 1/3$) as a result of some reheating process.  In the absence of any additional 
modifications of the expansion history, the universe would have remained RD until the 
time $t_{\rm MRE}$ of matter-radiation equality (corresponding to a redshift around 
$z \sim 3400$), transitioned to an epoch of matter domination (with $w = 0$), and remained 
matter-dominated (MD) so until very recent times (corresponding to redshifts 
$z \lesssim 2$), at which point dark energy came to represent a significant fraction 
of the energy density of the universe.  Since this is the sequence of cosmological epochs 
which follows from what may in some sense be regarded as the ``minimal'' model which addresses 
the flatness and horizon problems while also yielding a universe which is consistent with BBN 
and CMB data, this sequence of epochs may be regarded as the ``standard expansion history.''

There are, however, compelling motivations for considering cosmological scenarios 
which give rise to {\it modified}\/ expansion histories --- \ie, expansion histories 
which differ from this standard picture.  First of all, data anomalies and tensions 
(\eg, the Hubble~\cite{Verde:2019ivm} and $S_8$~\cite{Battye:2014qga} tensions) 
have recently emerged between the values of 
cosmological parameters at high redshifts $z \sim 1100$ (corresponding to the time of
CMB decoupling) and measurements at much lower redshifts $z \lesssim 5$. 
Scenarios with modified expansion histories can provide ways of addressing these tensions.
Moreover, modified expansion histories arise naturally --- and in some cases 
almost inevitably --- in many top-down scenarios for new physics~\cite{Kane:2015jia}.  In 
many cases, this is a consequence of the additional particle content present in such models.  
Theories with extra spacetime dimensions, for example, give rise to towers of 
Kaluza-Klein (KK) resonances.  String theory not only gives rise to such towers, 
but also to moduli and string axions.  Such particles can come to dominate the 
energy density of the universe.  A number of mechanisms, including the collapse 
of primordial overdensities after inflation or the collision of bubble walls 
during a first-order phase transition, can give rise to a non-negligible population 
of primordial black holes (PBHs).  These PBHs can subsequently come to dominate
the energy density of the universe.

Such modifications of the cosmological expansion history can have a variety of 
consequences for observation.  First and foremost, the manner in which perturbations 
evolve depends on the effective equation-of-state parameter for the universe as a whole.  
As a result, the manner in which this equation-of-state parameter has evolved over
time has implications for gravitational-wave (GW) production and 
evolution~\cite{Giovannini:1998bp,Giovannini:1999bh,Boyle:2005se,Watanabe:2006qe,
Boyle:2007zx,Kuroyanagi:2008ye,Kuroyanagi:2010mm,Kuroyanagi:2014qza,Figueroa:2019paj}.  
The manner in which perturbations in the dark-matter (DM) density evolve over time likewise 
depends on this equation of state parameter~\cite{Erickcek:2011us,Barenboim:2013gya,
Fan:2014zua,Redmond:2018xty}.  As a result, modifications to the expansion
history during or subsequent to the epoch wherein the DM is produced can have 
implications for small-scale structure, including potentially the formation of novel 
compact objects such as microhalos~\cite{Erickcek:2011us,Erickcek:2015jza,Delos:2023vfv} 
and even in the presence of multi-component DM~\cite{WileyDeal:2023trg}.  
The equation of state of the universe during the period which immediately follows inflation 
is also intimately related to the impact that inflaton self-resonance 
phenomena~\cite{Lozanov:2017hjm} (including, \eg, oscillon production~\cite{Amin:2011hj}) 
can have on the subsequent cosmological dynamics.

There are also numerous possible ramifications for DM physics.
For example, many scenarios which lead to modifications of the expansion history --- such
as those in which a population of heavy, unstable particles or PBHs gives rise to 
an early matter-dominated era (EMDE) or period of cosmological stasis --- can also lead 
to the non-thermal production of DM~\cite{Moroi:1999zb,Fujii:2002kr,Gelmini:2006pw,
Acharya:2008bk,Acharya:2009zt}.  Modified time-temperature relations and other 
consequences of modifications to the expansion history can impact the production 
of DM via mechanisms such as freeze-out or freeze-in~\cite{Co:2015pka,Evans:2016zau,Redmond:2017tja}.  
The DM relic abundance can also be modified after the DM is produced (\eg, via entropy dilution 
at the end of an EMDE)~\cite{Scherrer:1984fd,Asaka:2006ek,Patwardhan:2015kga,Randall:2015xza,
Berlin:2016vnh,Tenkanen:2016jic,Berlin:2016gtr,Hamdan:2017psw,Hardy:2018bph,Bernal:2018ins,
Chanda:2019xyl}. 

There are also potential ramifications for baryogenesis.  For example, entropy dilution 
at the end of an EMDE not only impacts the DM abundance, but can also wash out a significant 
portion of a previously generated baryon asymmetry~\cite{Kane:2011ih}.  This effect can 
be a problem or an asset, 
depending on circumstances.  In addition, in scenarios involving a significant population of
PBHs, a baryon asymmetry can in principle be generated if the particle species into which the
PBHs can evaporate include heavy, CP-violating particles which can subsequently generate a
baryon asymmetry via their decays~\cite{Toussaint:1978br,Turner:1979zj,Barrow:1990he,
Baumann:2007yr,Fujita:2014hha,Morrison:2018xla,Hooper:2020otu,Perez-Gonzalez:2020vnz,
Bernal:2022pue,ShamsEsHaghi:2022azq}.  In baryogenesis scenarios which take advantage of this 
mechanism, the cosmological epoch during which the PBHs are produced can have a significant 
effect on the resulting baryon asymmetry.

We see, then, that there are many intriguing possibilities --- both theoretical and 
observational --- for non-standard cosmological epochs and expansion histories.  The 
conversations recorded in the sections that follow take up many of these ideas in further 
depth.


\section{Block~I: Connection to Observation: Overview\label{sec:block1}}


\begin{center}
{\bf Discussion Leaders}\/: \\ Andrew J.~Long and Kuver Sinha
\end{center}

Since this workshop is largely about ``non-standard'' cosmological expansion histories, 
perhaps we should start by discussing what the ``standard'' cosmology is.  There is a 
lot of room for debate here, particularly with regard to issues like baryogenesis, the 
origin of DM, early accelerated expansion (\ie, inflation), and the nature of 
dark energy.  Everyone is welcome to have their own definition for ``standard.''  
However, I think we might all agree that any reasonable definition of standard cosmology 
would incorporate radiation-domination at a plasma temperature 
$T \sim (\mbox{a few}) \; \mathrm{MeV}$, which allows for neutrino 
thermalization/freeze-out and BBN, followed by cosmological expansion and adiabatic 
cooling leading to a transition to DM domination at $t_{\rm MRE}$, baryon 
acoustic oscillations, recombination, the ``dark ages,'' reionization, and the rest of 
late-time cosmology.  

Where is there room for something ``non-standard?''  During or after matter-radiation 
equality (MRE), modifications to the expansion history are strongly constrained by a 
variety of observational probes.  The late-time expansion rate (\ie, the Hubble constant 
$H_0$) is inferred from distance-ladder measurements --- for example using Cepheid-variable 
stars and Type-Ia supernovae out to redshift $z \approx 1$.  The expansion history at yet 
earlier times is constrained by observations of the anisotropies in the CMB radiation as 
well as the spatial distribution of matter/galaxies on 
cosmological scales, which both bear the imprints of baryon acoustic oscillations (BAO) 
that took place in the early universe.  The spatial distribution of DM on smaller 
length scales is probed by various observations, including Lyman-$\alpha$ forest spectra. 
These measurements are frequently framed in terms of the linear (dark) matter power spectrum. 

\begin{itemize}

\item Question: How much ``wiggle room'' remains in the matter 
power spectrum?  More specifically, can you change different experimental issues/systematics 
that affect various parts of the matter power spectrum in different ways and still 
get an overall set of results that are consistent with data?  

\item Answer: Yes, in principle, but there are a lot of subtle issues that one would 
need to deal with.  The non-linear complications that need to be accounted for at all 
relevant scales (even those at low values of the wavenumber $k$, for which CMB data 
are the dominant observational handle) are model-dependent, and a $\Lambda$CDM cosmology
is assumed when people make standard plots.  

\end{itemize}

Between BBN and MRE there is some limited freedom to modify the expansion history 
without coming into conflict with observations.  An entropy injection would disrupt 
the excellent agreement between the observed light-element abundances and the predictions 
of BBN~\cite{Sobotka:2022vrr,Sobotka:2023bzr}.  However, if there is no entropy injection, 
then even an $\mathcal{O}(1)$ modification of the expansion history is still 
allowed if it occurs sufficiently long after BBN and before MRE.  Adrienne Erickcek 
and Tristan Smith have recently been working on some things along these 
lines~\cite{Sobotka:2024ixo}.

\begin{itemize}

\item Question:  Other than the Hubble and $\sigma_8$ tensions, what motivates modifications
of the expansion history {\it after}\/ BBN?  

\item Answer: Maybe dynamical models that address the cosmic-coincidence problem.

\end{itemize}

On the other hand, prior to BBN -- it's the ``wild west.''  To learn about the expansion 
history of the early universe, we must rely on the cosmological relics that survive from 
that time.  Broadly speaking, the expansion rate of the universe can impact a relic 
through either propagation or production.  

As an example of the expansion history impacting a relic through propagation effects, let's 
discuss primordial gravitational waves.
Quantum fluctuations in the metric are expected to give rise to a broad spectrum of 
primordial GWs.  For modes that re-enter the horizon during a RD era 
(\ie, one in which the equation-of-state parameter for the universe is $w=1/3$), this 
spectrum is predicted to be scale invariant (roughly equal energy per logarithmically-spaced 
frequency interval).  However in a non-standard expansion history, the spectrum is modified, 
developing a blue tilt for $w > 1/3$ and a red tilt for $w < 1/3$.  A detection of these 
high-frequency primordial GWs with interferometers like LIGO-Virgo-KAGRA or the future 
space-based interferometer LISA would prove invaluable. 

A second example of the expansion history impacting a relic through propagation effects is 
found in warm-dark-matter scenarios.  If the DM is produced with significant velocity 
dispersion, then it free streams an appreciable distance during the course of the cosmic history.  
This free streaming suppresses the growth of structure on smaller length scales, possibly leading 
to observable signatures.  Since the free-streaming-length calculation involves an integral over 
cosmic time, it is impacted by the expansion history.  For example, in the warm-wave dark-matter 
(WWDM) scenario~\cite{Amin:2022nlh,Liu:2024pjg,Ling:2024qfv} the DM is ultralight 
(with a mass $m \sim 10^{-20} - 10^{-18}$~eV) and either relativistic or non-relativistic at the time 
of production, but nonrelativistic before MRE.  Depending on the WWDM mass and typical comoving 
momentum, the free-streaming length scale may be large enough to be probed with Lyman-$\alpha$ forest 
observations, and the limit depends on the expansion history.  

Cosmic expansion can also impact relics at the time of their production.  If DM freezes out 
at/around the electroweak phase transition (EWPT), there's a $\sim 1$\% effect on the expansion 
rate from the contribution to the total energy density of the universe that comes from the 
Higgs condensate.  This can actually impact the abundance of DM a bit if indeed DM freeze-out 
happens right around the EWPT~\cite{Chung:2011hv,Chung:2011it}.  A larger effect is possible 
in beyond-the-Standard-Model (BSM) scenarios with a first-order EWPT and strong supercooling.

\begin{itemize}

\item Question: Can you do the same thing with the quantum-chromodynamics (QCD) phase 
transition?  

\item Answer: Probably, but we're not aware of any work on it.

\end{itemize}

So what exactly do we mean by non-standard cosmological expansion histories?  To provide 
an example of some of the gray areas involved in delineating standard from non-standard 
histories here, I'll point out that high-scale inflation (meaning that 
the energy scale during inflation is GUT-scale or higher) used to be regarded as ``standard.''  
However, it might not be regarded as such anymore because we haven't seen evidence in the CMB of 
the GWs that such high-scale inflation would typically generate in single-field models.  
However, in multi-field inflation, high-scale inflation might still be ``standard.''  

\begin{itemize}

\item Question: Where in the ``First Three Seconds'' 
review~\cite{Allahverdi:2020bys} do observational signals come in?  It is a challenging 
question and not a lot is covered there on this topic.  How {\it do}\/ we test or exclude 
possibilities like these?  

\item Answer: We have GWs.  We have modifications to the properties of DM halos such as 
enhanced DM substructure and so forth.  

\item Comment: When DM particles are produced non-thermally from decaying 
fields at the end of an EMDE or during a stasis epoch, they often don't kinetically 
equilibrate with each other or with DM particles previously produced by other means.  
What does this do to the phase-space distribution $f(p)$ for DM?  What are the 
implications for free-streaming?  Can one extract information on this from the matter power
spectrum?

\item Question: Is it possible as a community to make a push toward gaining the ability to take 
an arbitrary set of density perturbations and other inputs and then generate power spectra 
and other outputs that account for non-linear effects?  

\item Answer: That would be hard!  Would we even know the full phase-space distribution of 
the DM?  The full $f(p)$ is probably hard to determine, but even if we can't do that, 
perhaps we could think about what useful characteristics of $f(p)$ we could in fact determine.  

\item Comment: Maybe we could obtain statistical ``moments'' of $f(p)$ 
(such as its average, width, skewness, and other higher moments).  Or maybe 
that's not the right strategy, but there could be other, more useful decompositions or 
parametrizations.  

\item Comment: A lot of parametrizations which are in use now are motivated by ultraviolet (UV) 
physics.  EMDEs in many ways followed from understanding that there were moduli in UV theories.  

\item Comment: It might be nice to talk more as a field in terms of quantities
that are more closely related to things that are measured/observed directly.

\item Comment: An effective-field-theory (EFT) approach in cosmology can also be 
useful for understanding the (p)reheating process~\cite{Ozsoy:2015rna,Ozsoy:2017mqc}.

\item Comment: Needing to run $N$-body and hydrodynamic simulations is a time bottleneck 
for properly fitting new physics with small-scale observations.  It's just not reasonable 
to make hosts of simulation suites for each new-physics scenario on the market, varying 
particle-model and standard cosmological parameters.  Simulations are getting faster 
(GPUs are an asset), and emulators help for running a Markov chain Monte Carlo (MCMC)
simulation, but the practicality and feasibility is still a concern.

\item Comment: An analogy with BSM searches at colliders may be useful. 
Back in the day, the first supersymmetry (SUSY) constraints were depicted within the 
$m_0 - m_{1/2}$ plane of the restricted paramter space of minimal supergravity (mSUGRA).
This analysis framework was followed by the EFT framework and then the simplified-model 
framework for interpreting results from both the Large Hadron Collider and DM searches. 
Would something akin to that kind of parametrization be useful for non-standard cosmologies, 
in terms of broadening our analyses beyond simply focusing on benchmark values for the equation 
of state (say, those associated with RD, MD, and kination-dominated cosmologies) and into 
a more general framework?

\end{itemize}


\section{Block~II: Scalar Fields and Non-Standard Expansion Histories~I 
(Top-Down Realizations and Implications for Structure Formation)
\label{sec:block2}}


\begin{center}
{\bf Discussion Leaders}\/: \\ Rouzbeh Allahverdi and M.~Sten Delos
\end{center}

\begin{figure}[t]
\begin{center}
\includegraphics*[width=3.0in]{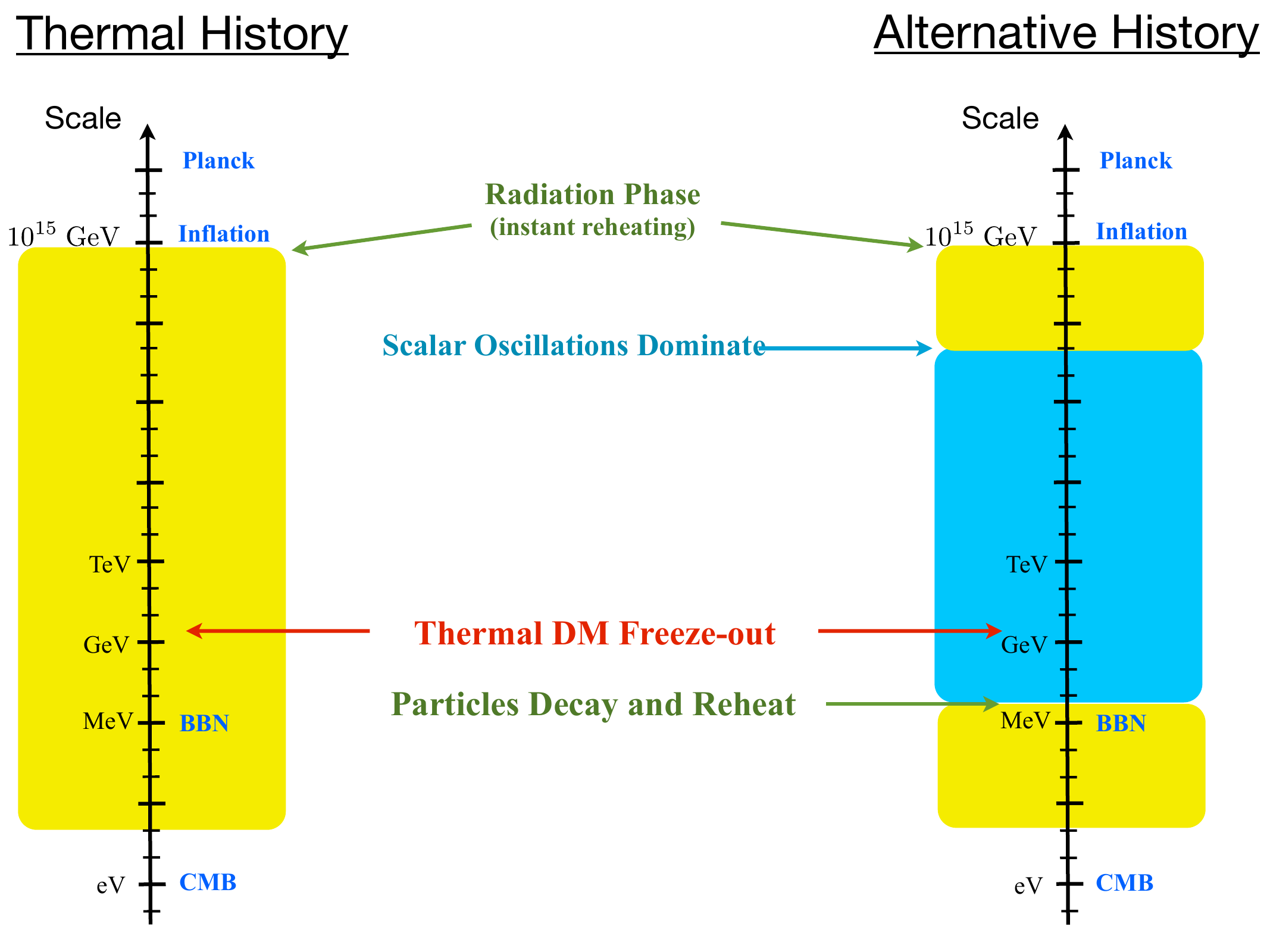} 
\hspace{0.2cm}
\end{center}
\caption{The timeline on the left represents a thermal history for the early 
universe wherein a population of WIMPs which constitute the DM is produced in the thermal 
bath that emerges during reheating, shortly after after inflation.  The timeline on the right 
represents a non-thermal history wherein the universe is dominated by heavy moduli (which 
behave like massive matter) up until the time at which the moduli decay --- a time which must 
occur before BBN.  Figure taken from Ref.~\cite{Kane:2015jia}.
\label{fig:ModuliDomTimeline}}
\end{figure}

This discussion block will focus on top-down realizations of scalar-field cosmologies.  
The coherent oscillations of a scalar $\phi$ around its minimum in a potential of the
form $V(\phi) \propto |\phi|^n$ yield a time-averaged equation-of-state 
parameter~\cite{Turner:1983he}
\begin{equation}
  \langle w\rangle ~=~ \frac{n-2}{n+2}~,
\end{equation}
so $n > 4$ gives us $1/3 < w < 1$.  In scalar-field scenarios of this kind, one can obtain a sort of 
``reheating'' without entropy generation.  Particular cases of interest are those involving epochs of 
kination ($w=1$) or early dark energy ($w < -1/3$).  What happens is that the coherent state of $\phi$ 
in which this field oscillates around the minimum of $V(\phi)$ fragments over time.  This fragmentation 
can give rise to $Q$-balls~\cite{Cotner:2019ykd} and 
$I$-balls/oscillons~\cite{Amin:2010xe,Amin:2010jq,Amin:2011hj,Gleiser:2011xj,
Hong:2017ooe,Fukunaga:2019unq} --- plus there's also self-resonant production of inflaton 
quanta~\cite{Lozanov:2017hjm}.  Thus, coherent oscillations don't last forever.

If the scalar $\phi$ interacts with some other scalar $\chi$, you can get rapid 
energy transfer from the coherent oscillations of $\phi$ to field quanta of $\chi$.
As an example, let's consider the scalar potential
\begin{equation}
  V(\phi,\chi) ~=~ \frac{1}{2} m^2\phi^2 + \frac{1}{2} g\phi^2\chi^2 
    + \frac{1}{2} \sigma \phi \chi^2 + \frac{1}{4}\lambda \phi^4~,
  \label{eq:VphiScalPot}
\end{equation}
where $\lambda$ and $g$ are dimensionless couplings, where $m$ and $\sigma$ have dimensions of mass,
and where $\chi$ is really light, so it's mass can be ignored.  This potential gives rise 
to such phenomena.  For this form of the potential, you get a plasma of $\phi$ and $\chi$ particles 
in equilibrium.  This plasma is generated either via parametric resonance~\cite{Podolsky:2005bw} 
(when the $\phi^2 \chi^2$ term dominates) or via tachyonic resonance~\cite{Dufaux:2006ee} 
(when the $\phi \chi^2$ term dominates).  As a result, the equation-of-state parameter for the 
universe remains around $w \approx 1/4$ after the generation of this plasma for an extended period 
of time. 

Another approach would be to use the EFT of (p)reheating~\cite{Ozsoy:2017mqc,Ozsoy:2015rna}. 
The ideas that underlie this approach stem from strategies that were previously developed 
in a particle-phenomenology context for dealing with searches for BSM physics --- searches 
wherein the fundamental underlying theory is unknown.  This approach has been useful when 
extended to cosmology in several ways.  These include the EFT of Inflation~\cite{Cheung:2007st}, 
the EFT of large-scale structure~\cite{Baumann:2010tm}, and the EFT of dark
energy~\cite{Gubitosi:2012hu,Bloomfield:2012ff}.

In other words, given our lack of knowledge about what drives the (p)reheating process, 
one can focus on symmetry principles and universality classes and treat $\phi$ 
not necessarily as a fundamental scalar, but as an order parameter associated with the
symmetry-breaking transition~\cite{Barrowes:2024akq}.  The goal of this approach is to 
establish a {\it model-independent}\/ approach to (p)reheating.  This is motivated by
the fact that much of the community has communicated dismay at how model-dependent 
(p)reheating is -- this was a topic of our discussions in advance of this
workshop.  It is important to note that potential in Eq.~(\ref{eq:VphiScalPot}) is an 
interesting example, but in general one must also worry about back-reaction effects 
and turbulence.  These are challenges for the current EFT approach.  However, it would be 
interesting to apply techniques similar to those used in the EFT of inflation to account 
for such effects~\cite{LopezNacir:2011kk}. 

So the question, then, is whether one can build a model that solves problems like 
the DM problem, the matter/anti-matter asymmetry problem, \etc, with moduli fields.  
The goal is to accommodate the correct DM relic abundance, to explain BAU, and not to 
overproduce dangerous relics (such as gravitinos) or dark radiation. The need to satisfy 
all of these requirements may have non-trivial implications for UV-complete models with 
non-standard thermal histories (\eg, models which emerge in the context of string
theory~\cite{Allahverdi:2013noa,Allahverdi:2014ppa,Allahverdi:2016yws,Allahverdi:2020uax}).  

\begin{itemize}

\item Question: what's the difference between Kachru-Kallosh-Linde-Trivedi
(KKLT)~\cite{Kachru:2003aw} and large-volume-scenario 
(LVS)~\cite{Balasubramanian:2005zx} scenarios?  

\item Answer: Both KKLT ans LVS start with anti-de Sitter (AdS) space and ``uplift'' the geometry.  
LVS breaks SUSY.  KKLT preserves SUSY.  Both of these can affect CMB observables --- for 
example, they have an effect on the tensor-to-scalar ratio $r$ and the spectral index $n_s$.  

\item Question: Can one isolate the effect of an EMDE (or multiple EMDEs) on $r$ and 
$n_s$?  

\item Answer: Yes, in principle, one could.  Any history that includes a epoch (or multiple epochs) 
with $w < 1/3$ results in a smaller number of $e$-folds being accessible in CMB observations.  
This in turn affects the scalar spectral index $n_s$ in a quantifiable manner within different 
universality classes of inflationary models~\cite{Roest:2013fha}.  This can be used to constrain 
the parameters of explicit models of the early universe that incorporate both inflation and 
the post-inflationary evolution of the universe. 

\item Question: Are there phenomena other than coherent oscillations of a scalar
that can give $w > 1/3$?  What is the fate of coherent oscillations?  

\item Question: Are there correlations between parameters related to 
reheating/post-inflationary cosmology and those related to inflation?  

\item Answer: We don't know.  The key challenge is to establish a UV-complete 
theory of both inflation and (p)reheating.  In the context of string-theory approaches, such
a UV-complete theory has been difficult to realize thus far.

\item Question: Are networks of topological defects a generic outcome from 
these string-inspired models?

\item Answer: Such networks {\it can}\/ form, but there's nothing really generic that can be
said along these lines.  Even whether the universe ends up being RD, MD, or neither of the above 
after inflation isn't generic.

\end{itemize}

Now let's take a look at the consequences of non-standard expansion histories in
scenarios with scalar-dominated epochs.  We focus on their impact on the small-scale 
distribution of the DM, and we consider scalars that do not themselves gravitationally 
cluster, so these scalar-dominated epochs influence the growth of DM density 
perturbations only by altering how DM particles drift over time.  In the absence of 
peculiar gravitational sources, nonrelativistic particles drift a comoving distance $|\vecbf{s}|$ 
proportional to 
\begin{equation}\label{eq:drift}
  |\vecbf{s}| ~\propto~ \int \frac{dt}{a^2} ~\propto~ \int \frac{da}{a^3 H}~,    
\end{equation}
where $a$ is the scale factor.  However, perturbations to the DM density [as described by the
the density contrast $\delta_{\rm DM}(\vecbf{x})$] are simply given by 
$\delta_{\rm DM}(\vecbf{x}) = -\vecbf{\nabla}\cdot\vecbf{s}(\vecbf{x})$, where $\vecbf{s}(\vecbf{x})$ 
is the field of particle displacements (as a function of the comoving position coordinate $\vecbf{x}$) 
from their initial (or ``Lagrangian'') positions.  Consequently, if the universe has the 
equation-of-state parameter $w$, matter perturbations grow as
\begin{equation}
  \delta_{\rm DM} ~\propto~ \int \frac{da}{a^3 H}~\propto~ 
  \begin{cases} 
     a^{(3w-1)/2} & w~\neq~ 1/3 \\ 
      \log a &  w~=~1/3~. 
    \end{cases}
\end{equation}
For example, perturbation growth is more efficient during a kination epoch ($w=1$), wherein
$\delta\propto a$, than it is in a RD epoch ($w=1/3$), wherein $\delta\propto\log a$.

In order to understand the spectrum of density perturbations that arises after scalar 
domination, it is also necessary to consider when they start growing.  The Fourier transform
$\delta_k$ of $\delta_{\rm DM}(\vecbf{x})$ with comoving wavenumber $k$ begins to grow at horizon entry, 
when $k\simeq aH$, because the DM receives a gravitationally ``kick'' from  
inhomogeneities in the energy density of the dominant species (radiation in the standard cosmological 
scenario or a scalar field in certain modified cosmological scenarios).  These inhomogeneities are 
short-lived after horizon entry, since the corresponding perturbation mode in the density contrast for
the dominant cosmological energy component likewise becomes dynamical at this point, but the motion of 
the DM persists and is responsible for the growth of perturbations prior to $t_{\rm MRE}$.  
However, the time at which a given mode enters the horizon is affected by the equation of state of 
the universe, since $aH\propto a^{-(3w+1)/2}$.  Putting these considerations concerning horizon-entry 
timing and drift efficiency together, one can show that the DM power spectrum $\Delta^2(k)$ 
scales with $k$ like
\begin{equation}
    \Delta^2(k) ~\propto~ k^{2\left(\frac{3w-1}{3w+1}\right)}\mathcal{P}_\zeta(k)~,
\end{equation}
where $\mathcal{P}_\zeta(k)$ is the primordial curvature power spectrum. For a kination epoch, 
this implies a fairly gentle increase $\Delta^2\propto k$ in the power spectrum with $k$ on scales 
that were within the horizon during the kination epoch, as pointed out in 
Refs.~\cite{Redmond:2018xty,Delos:2023vfv}.

For power spectra that rise on small scales, it is important to understand where that rise 
stops --- \ie, where the power spectrum cuts off on small scales.  This cutoff is typically driven 
by random thermal motion of the DM, which erases density perturbations on scales smaller 
than the distance that particles randomly stream.  This ``free streaming'' scale is affected by the 
efficiency of particle drift [\cf\ Eq.~(\ref{eq:drift})], and it is also affected by when and at 
what temperature particles begin to stream.  For DM that kinetically decouples from 
radiation during a kination epoch, it turns out that the free streaming scale is smaller than 
if the same DM decoupled during radiation domination. This means that the matter power 
spectrum is preserved up to higher $k$.

If the DM power spectrum has a higher amplitude on some scale $k$, as would arise 
after a kination epoch, this means that bound, virialized DM halos begin to form 
at earlier times.  This is because virialized systems form when density perturbations grow to 
be of order $\delta \sim 1$.  However, earlier-forming halos are denser, with an internal density 
which is proportional to the density of the universe at their formation time.  Consequently, 
cosmological dynamics that lead to amplified power can potentially leave detectable signatures 
in the small-scale spatial distribution of the DM.

The power spectrum of the DM has been measured for wavenumbers as high as $k \sim 50$~Mpc$^{-1}$, 
corresponding to $\sim 10^7$~M$_\odot$ mass scales~\cite{Gilman:2021gkj,Sabti:2021unj,Esteban:2023xpk}.
Probes of smaller mass scales exist, but they can only constrain scenarios in which the power 
spectrum is greatly amplified.  For example, mass scales around $M \sim (10$ -- $10^6)$~M$_\odot$ could 
be probed by looking at how small halos gravitationally heat systems of stars, such as dwarf 
galaxies~\cite{Graham:2024hah} or stellar binaries~\cite{Ramirez:2022mys}.  Still smaller mass 
scales, as low as $M \sim 10^{-10}$~M$_\odot$, can be constrained by searching for how DM 
microhalos gravitationally lens light from stars~\cite{Delos:2023fpm,Dai:2019lud,Blinov:2021axd} 
or perturb the motion of pulsars~\cite{Ramani:2020hdo,Lee:2020wfn,Delos:2021rqs}.  Smaller, 
asteroid-level mass scales could lead to perturbations in the motions of Solar System bodies or spacecraft.

Another phenomenological effect of small, compact DM halos is the effect on DM annihilation.
If the DM annihilates, the annihilation rate is dominated by the contribution from the 
smallest, densest structures~\cite{Delos:2022bhp}.  DM that freezes out during a kination 
epoch is tightly constrained, partly for this reason~\cite{Delos:2023vfv} and partly because 
DM which freezes out during kination must have a higher annihilation cross-section to 
reach the known DM abundance~\cite{Pallis:2005hm}.

\begin{itemize}

\item Question: What fraction of the DM ends up in these smallest 
structures?  

\item Answer: It varies, depending on the cosmology.  For a standard power spectrum, arising from 
scale-invariant inflationary perturbations and a standard expansion history, they typically 
contain around $1$\% of the DM~\cite{Delos:2022bhp}.  If the power spectrum is boosted at small 
scales, that figure can be of order $10$\% or higher.

\item Comment: The energy density associated with a kination component
gets suppressed automatically as time increases because it has $w > 1/3$, so you don't 
need a ``graceful exit'' to make it to go away.  

\item Question: Do instanton corrections mess up kination?  Don't they 
generate a mass for the field?  

\item Answer: Yes, in general, but the mass can be small, so you can still get 
$w \approx 1$ if not exactly $w =1$.  For example, if you have a direction in field space that's 
associated with a truly flat compact direction --- a pure Goldstone boson --- and the scalar field is 
rolling around that flat direction, that's pure kination.  If there's a tilt to the potential
associated with some explicit breaking of the symmetry like you'd get from instantons, 
now there's a minimum, but if the tilt is small the field can cycle around a bunch of times 
if its initial field velocity is large.  So yes, instanton-generated masses exist, but they 
can be really small and you can still achieve $w \approx 1$.  

\item Question: Does a field with $w < 0$ cluster --- say, as a result of 
fragmentation?  

\item Answer: If an energy component with $w < -1/3$ is dominating the energy 
density of the universe, there's accelerated expansion, so perturbations are diluted by inflation.  
For $-1/3 < w < 0$, however, it's not immediately clear.  

\item Question: What is the survival probability of the mini-halos as substructure within our galaxy? 
Is the fraction of DM within mini-halos very different within our galaxy, compared to the fraction 
generated cosmologically?

\item Answer: This generally depends on how dense the clusters are, because that determines how 
resistant they are to the mass loss that can result from tidal forces and encounters with stars. 
However, even in an extreme scenario where most of the matter is in mini-clusters that formed 
a moderate length of time before MRE (and thus are about as dense as possible), only around $10$\% 
of the DM at our position within the Milky Way would be in these clusters~\cite{OHare:2023rtm}.  
In less extreme scenarios, the amount of mass loss would be greater.

\end{itemize}


\section{Block~III: Gravitational-Wave Signals in Modified Cosmologies\label{sec:block3}}


\begin{center}
{\bf Discussion Leaders}\/: \\ Kimberly K.~Boddy and Lauren Pearce
\end{center}
    
A contribution to the stochastic GW background is know to arise at second order from 
first-order curvature perturbations~\cite{Ananda:2006af}.  Even though these 
perturbations arise at second order, they can still yield large enhancements.  This is 
related to the so-called poltergeist effect~\cite{Inomata:2020lmk}.  The origin of this 
enhancement may ultimately be a resonance effect: an oscillating gravitational potential 
yields an oscillating sound speed~\cite{Ananda:2006af,Inomata:2020lmk}.  The 
resulting enhancement in the GW spectrum occurs over a particular range of wavenumbers
$k$ during a MD epoch, and both this range of $k$ and the extent of the enhancement 
depend on how the MD epoch ends~\cite{Inomata:2019ivs,Inomata:2019zqy,Pearce:2023kxp}.  
If the transition from the MD epoch to the epoch that follows it is slow (and ``slow,'' 
according to this definition, includes the standard exponential fall-off associated with 
particle decay), there's no enhancement, and in fact, careful calculation shows that the 
signal is suppressed due to a cancellation between terms~\cite{Inomata:2019zqy}.  
By contrast, when the MD era ends more rapidly, there {\it is}\/ a significant 
resonance-like enhancement~\cite{Inomata:2019ivs}.  The analysis of the relevant
dynamics was initially performed in the unphysical limit of an instantaneous transition between 
matter and radiation domination, but recent work interpolates between the two 
regimes~\cite{Pearce:2023kxp}, clearly showing the interplay between the resonance and the 
suppression.

So how do we get such a fast phase transition (from early matter domination to radiation 
domination)?  We need heavy objects whose decays speed up.  In principle, 
PBHs~\cite{Inomata:2019ivs} or $Q$-balls~\cite{White:2021hwi} would work.

The signal increases the longer the EMDE lasts, as perturbations 
have a longer period to grow.  However, analyses are cut off at the scale where the 
perturbations start to become non-linear.  This leads to the sharp cutoff in the signal 
for $k$ above some $k_{\rm max}$ set by the non-linear cutoff.  Really, we should 
consider the GW signal to be unknown in this region.

Work has also been done to calculate the contribution to the GW spectrum which arises
at second order in the presence of sizable curvature perturbations when the 
power spectrum is modified.  Such perturbations can be generated via a phase transition, 
which can also modify the evolution of the scale factor~\cite{Lewicki:2024ghw}.

Moreover, we can also consider alternative early-universe cosmologies in which the early MD epoch is 
replaced by an epoch with a different value of $w$.  This has been done in some generality 
(although with less attention paid to what happens at the end of the epoch)~\cite{Domenech:2020kqm}, 
and for kination, resulting GW the signal can lie within the frequency range of interest for 
pulsar-timing arrays (PTAs)~\cite{Harigaya:2023pmw}.

In particular, PTAs are sensitive to GWs in the nHz frequency regime.  Let's consider sources 
for GWs in this range.  One (standard) source is merging supermassive-black-hole (BH) binaries.  In 
the early universe, we can also have GWs produced during inflation, GWs produced during phase 
transitions, and GWs produced by topological defects in this frequency range.  From an observational
perspective, we now have evidence for a stochastic GW background within this frequency range from 
NANOGrav~\cite{NANOGrav:2023gor}, EPTA/InPTA~\cite{EPTA:2023fyk}, PPTA~\cite{Reardon:2023gzh}, 
and CPTA~\cite{Xu:2023wog}. 

\begin{itemize}

\item Question: Does NANOGrav measure a three-point or four-point 
correlation function for the GW background?  

\item Answer: PTAs have reported evidence for the monopole component of the GW background. 
The Hellings-Downs curve~\cite{Hellings:1983fr} represents the spatial correlations of the 
timing residuals between pairs of pulsars for an isotropic GW background.  We do not have a 
sky map of the GW anisotropies.  Moreover, at the lowest frequencies, where the signal is 
largest, the GW background is expected to be Gaussian, in which case, higher-point correlation 
functions would not provide new information.

\item Question: Can one construct a curve similar to the Hellings-Downs curve for sets of three 
given pulsars?  For sets of four pulsars?  

\item Answer: It is not clear what the advantage of this would be.  Correlating pulsar pairs 
helps distinguish the GW signal from sources of noise and systemics.  I am not aware of any 
work that considers how correlating sets of three or more pulsars would be helpful.

\item Question: Why is NANOGrav sensitive to frequencies in the nHz range in 
particular?  What sets the frequency scale of PTAs?  

\item Answer: The lower limit on the frequency $f$ to which a PTA experiment is sensitive 
is set by lifetime of the observing mission.  The upper limit is set by the cadence of 
observation.  The stochastic GW background has a red-tilted spectrum, so the data at lower 
frequencies has a higher signal-to-noise ratio.  Although it would not be ideal for measuring 
the GW background, higher frequencies could be accessed with higher cadence observations, aided 
by adding more telescopes.

\end{itemize}

As far as new-physics interpretations of GW signals at PTAs are concerned, as we've seen, 
it's really only the lowest couple of frequency bins that matter.  Astrophysical sources 
(supermassive BHs) fit the current NANOGrav data well~\cite{NANOGrav:2023hfp}, given uncertainties 
in population-ellipticity statistics and environmental issues.  For example, BHs are in galaxies, 
and there are other energy-loss mechanisms~\cite{Yu:2001xp,Escala:2004jh,Dotti:2006ef,Haiman:2009te} 
for BHs in addition to GW emission in those environments.  

\begin{itemize}

\item Question: Can non-standard expansion histories (in the early universe) affect 
the abundance and distribution of masses for supermassive-BH binaries?  

\item Answer: No, binaries form from the mergers of galaxies hosting supermassive BHs, and these 
mergers occur at later times in the universe's history.

\item Question: How much better can/will the data get before we're 
all six feet underground?  

\item Answer: More pulsars and more data will shrink the current error bars and push towards 
lower frequencies.  The PTA collaborations actively search for new millisecond pulsars to add 
to the catalogue.  However, since the lowest frequency is set by the timescale of observation, 
we are limited in terms of how low a frequencies we can access on reasonable human timescales.  

\item Question: How can we do better than current PTA observations within this frequency range?  

\item Answer: We can search for anisotropies in the GW background.  This might allow us to 
distinguish cosmological from astrophysical sources of GWs.  The latter have much larger 
anisotropies than the former.  Note that an individual supermassive-BH-binary source may be 
detectable as a continuous wave source if its amplitude is sufficiently large, and would be 
treated as a foreground.  The astrophysical GW background arises from all the unresolved 
binary sources.

\item Question: Is there information about the polarization of the GW background 
signal?  

\item Answer: Yes.  While we cannot extract polarization information from the monopole, effects 
are present in the anisotropy of the background.

\item Question: Are there proposals for figuring out how to probe the frequency 
gap between LISA and SKA?  

\item Answer: Yes, there are.  First, there's a proposal for a space-based detector called 
$\mu$-ARES, which would be much like LISA in its design, but with gigantic distances between 
the detectors.  There are also other proposals for larger LISA-like instruments that wouldn't be 
quite as big as $\mu$-ARES would be.  There's also astrometry~\cite{Moore:2017ity}.  
One could use Gaia data~\cite{Gaia:2018ydn} --- and, in the future, Theia data~\cite{Theia:2017xtk}.

\item Question: Why can't we just fit a GW model to the gray ``violins'' 
on the NANOGrav plot, as shown in 
Fig.~\ref{fig:NANOGravViolins}?

\item Answer: These ``violins'' give us some sense as to the frequency power spectrum of the GW
background, but they do not include information about the angular correlations.

\begin{figure}
  \centering
  \includegraphics[width=0.5\textwidth]{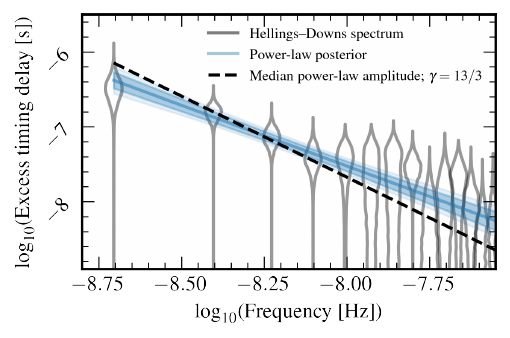}
  \caption{The results of a Bayesian ``free-spectrum'' analysis of the 
    NANOGrav 15-year data set, taken from Ref.~\cite{NANOGrav:2023gor}.  The gray 
    ``violins'' represent the posteriors of the frequency power spectrum at a discrete set 
    of frequencies $f_\ell = \ell/T$ (where $T$ is the total time span for the data set), 
    assuming a Hellings–Downs angular correlation.  The solid blue line represents 
    the posterior best fit to the data, while the black dashed line is the result
    of a fiducial model in which the GW background is generated by supermassive-BH 
    mergers.  For further details, see Ref.~\cite{NANOGrav:2023gor}.}
    \label{fig:NANOGravViolins}
\end{figure}

\end{itemize}


\section{Block~IV: Early Matter-Dominated Eras\label{sec:block4}}


\begin{center}
{\bf Discussion Leaders}\/: \\ Adrienne L.~Erickcek and Jessie Shelton
\end{center}

EMDEs are realized in many well-motivated theories, including those with moduli,
axions/ALPS, saxions, heavy particles that freeze out, \etc\  (see, \eg, 
Refs.~\cite{Kane:2015jia,Allahverdi:2020bys} for reviews).  From a phenomenological
perspective, they are characterized by their onset times and durations.  Generically, 
a sizeable entropy dump occurs at the end of an EMDE~\cite{Scherrer:1984fd}.  
The decay of an unstable particle species that dominates the energy density of the 
universe during an EMDE is ``gradual'' in the sense that the time window over which 
the majority of the individual particles decay is on the same order as the total 
time that the universe has existed prior to the time at which they decay.

An EMDE impacts the prospects for the terrestrial detection of DM in a number of ways.  
In general, the entropy dump at the end of the EMDE dilutes any preexisting DM abundance.  
This means that if DM is produced via thermal freeze-out prior to the end of the EMDE, a larger 
DM abundance needs to be produced than would otherwise have been necessary, which means 
that the couplings which govern the DM annihilation rate need to be weaker for the same DM
mass~\cite{Asaka:2006ek,Berlin:2016vnh,Tenkanen:2016jic,Berlin:2016gtr}.  Thus,
entropy dilution associated with the end of an EMDE can often open up a significant region 
of the parameter space of a given DM freeze-out model.  One simple 
example is furnished by a freeze-out scenario with a real scalar DM particle $S$ that 
couples to the SM via the Higgs-portal interaction~\cite{Hardy:2018bph} 
\begin{equation}
  \Delta \mathcal{L}_{\int} ~\ni~ -\frac{\lambda}{2}|H|^2 S^2~.  
\end{equation}
The modification of the Hubble-parameter evolution can also modify the freeze-out 
dynamics in cases in which the DM particle freezes out {\it during}\/ the 
EMDE~\cite{Hamdan:2017psw,StenDelos:2019xdk}.  In other words, the time at which the 
freeze-out criterion $n_{\rm DM}\langle\sigma v\rangle \sim H$ is achieved, where $H$ is the 
Hubble parameter and $n_{\rm DM}$ is the number density of DM particles, can be altered.  
Moreover, if the metastable particle species which dominates the energy density 
of the universe during the EMDE can decay to final states which include DM particles, 
these decays give rise to a non-thermal production channel in addition to the thermal 
one associated with freeze-out.  This effect can open up even more model-parameter space 
for freeze-out and make very small couplings phenomenologically viable.

By contrast, if the DM is produced by freeze-in prior to the end of an EMDE, 
{\it stronger}\/ couplings are needed to achieve the correct DM abundance than would otherwise 
have been needed in the absence of the EMDE.  Such couplings can potentially lead to 
signals at terrestrial experiments.  For example, let's consider a scenario in which some 
additional heavy particle species $B$ of mass $m_B$ couples to both the DM particle and to 
the visible sector via a single Lagrangian term which involves a $B$ particle, the DM particle, 
and some SM particle~\cite{Co:2015pka}.  Provided that $T \gtrsim m_B$ and that $B$ remains in 
thermal equilibrium with the radiation bath, DM particles can be produced 
by the interactions of the $B$ with other particles in the bath.  The couplings required in order 
to yield the correct DM abundance are such that this scenario could potentially give rise to striking 
signals at colliders.  Depending on the values of the model parameters, these signals could 
involve displaced vertices and/or stopped charged heavy particles (the $B$) within the detector.

If the DM is a light axion or ALP of mass $m$ produced via misalignment production, 
the late-time abundance obtained for this field can be modified due to the fact that 
the equation-of-state parameter of the universe is $w = 0$ rather than $w = 1/3$ during the
EMDE~\cite{Banks:1996ea,Blinov:2019rhb}.  This is because the time at which the criterion 
$3H \sim 2m$ is achieved and the field transitions from overdamped to underdamped oscillation 
(and thus from behaving like vacuum energy to behaving like massive matter) is altered by the 
modification of $w$.

Yet another interesting possibility one can consider is a hidden-sector DM scenario
in which the hidden sector consists of a fermion $\chi$, which plays the role of the DM, 
and another dark/hidden-sector particle species $\phi$ which can decay to SM particles 
and therefore plays the role of a mediator between the two sectors.  
In a scenario like this, $\chi\chi\to \phi\phi$ annihilation and other processes can produce 
metastable $\phi$ particles when the kinetic energy of the particles is large (\ie, at
early times).  The $\phi$ particles can dominate the universe before they decay, and this
EMDE can have observable consequences due to the effect on density 
perturbations~\cite{Zhang:2015era}.

The presence of an EMDE in the cosmological timeline can also have an impact on the matter 
power spectrum $\mathcal{P}(k)$. During an EMDE, perturbations in the DM density grow linearly 
with $a$ once they enter the horizon, so the power spectrum is enhanced for values of $k$ 
associated with modes that enter the horizon during or before the EMDE.  Therefore, the 
minimum $k$ at which perturbations are enhanced is the wavenumber 
$k_{\rm RH} = a_{\rm RH}H_{\rm RH}$ associated with the mode that enters the horizon just 
as the EMDE is ending and the universe transitions to RD.  There's also a high-$k$ cutoff 
$k_{\rm cut}$ above which $\mathcal{P}(k)$ is suppressed.  This can be due to the free streaming 
of the DM particles~\cite{Gelmini:2008sh,Fan:2014zua,Erickcek:2015bda,Waldstein:2016blt,Miller:2019pss}.  
On the other hand, if the DM is sufficiently cold, the value of $k_{\rm cut}$ typically depends on 
the properties of the metastable particle species that dominates the universe during the EMDE.  
For example, if this metastable species was once relativistic, it only begins to cluster like 
matter after its temperature $T_\mathrm{ms}$ falls well below its mass $m_\mathrm{ms}$. As a 
result, density perturbations on scales that enter the horizon while 
$T_\mathrm{ms} \gtrsim m_\mathrm{ms}$ are suppressed, and $k_{\rm cut}$ is about 0.1 times 
the wavenumber of the mode that enters the horizon when 
$T_\mathrm{ms} = m_\mathrm{ms}$~\cite{Ganjoo:2022rhk}.  Cannibalistic DM 
self-interactions~\cite{Dolgov:1980uu,Carlson:1992fn} can provide an additional source of 
pressure for the metastable species, leading a lower-$k$ cutoff in 
$\mathcal{P}(k)$~\cite{Erickcek:2020wzd,Erickcek:2021fsu}.

Another interesting effect arises if the density contrast becomes so large that 
$\delta_k \gtrsim 1$ during the EMDE for some range of $k$ .  In such cases, the metastable 
particles that dominate the energy density during the EMDE can form halos {\it during}\/ the 
EMDE.  DM particles fall into these halos and are released when the metastable particles 
decay and the halos disappear.  Since the DM particles are accelerated when they fall into the 
halos and are released in random directions, the formation of structure during the EMDE 
gravitationally heats the DM~\cite{Blanco:2019eij, Barenboim:2021swl, Ganjoo:2023fgg}. 
The subsequent free streaming of DM particles sets a new cutoff scale in the matter power 
spectrum~\cite{Ganjoo:2023fgg}.  There's also a contribution to the GW background from this 
process~\cite{Jedamzik:2010hq,Fernandez:2023ddy}.  

{\it After}\/ the EMDE, a significant fraction of the DM becomes bound in microhalos at very 
early times. For $k_\mathrm{cut}/k_\mathrm{RH}\gtrsim 40$, over half of the DM is 
contained in microhalos at a redshift $z \sim 400$~\cite{Erickcek:2015bda}.  By contrast, a 
significant population of DM halos never forms at redshifts higher than $z \sim 70$ -- $80$ in 
a universe with a standard expansion history. An EMDE can even lead to halo formation prior to 
MRE, during the RD epoch~\cite{Blanco:2019eij}!  These early-forming microhalos have a huge impact 
on the DM annihilation rate~\cite{Erickcek:2015jza,Blanco:2019eij,StenDelos:2019xdk,Ganjoo:2024hpn}.

The leading constraint on these halos is currently the indirect-detection bound from observations 
of the isotropic gamma-ray background~\cite{Blanco:2019eij,StenDelos:2019xdk,Ganjoo:2024hpn}. 
However, observations of the 21cm power spectrum might provide stronger constraints in the future.
DM annihilation in microhalos mimics DM decay because it depends on the number density $n_{\rm mh}$ 
of microhalos, which means the annihilation rate is proportional to the average DM
density $\overline{\rho}_{\rm DM}$, as opposed to $\overline{\rho}_{\rm DM}^2$.  One possible way to 
distinguish emission from annihilation within microhalos from emission from DM decay is to look 
at the gamma-ray emission profiles of dwarf spheroidals.  Microhalos are less likely to survive 
in the centers of galaxies (including dwarf galaxies), so the signal from annihilation within 
microhalos would be suppressed relative to the emission signal from decaying DM within the 
central region of the galaxy where $n_{\rm mh}$ is no longer proportional to 
$\overline{\rho}_{\rm DM}$~\cite{StenDelos:2019xdk}.  

\begin{itemize}

\item Question: What about CMB constraints?  Are they competitive with these
indirect-detection ones?  

\item Answer: No, they're not.  Microhalo formation changes the manner in which the annihilation 
rate scales with the scale factor from scaling as $a^{-6}$ (\ie, as the square of the average 
DM density) to scaling as $a^{-3}$ (\ie, as the number density of microhalos).  This slower 
decrease of the annihilation rate enhances the constraining power of present-day probes 
compared to earlier probes such as the CMB.  The CMB constraints on decaying DM are weaker 
than the constraints from the gamma-ray background for the same reason.

\item Question: Does the annihilation channel affect things in terms of
the qualitative results?  

\item Answer: A little, but not a lot.  For example, it is shown in Ref.~\cite{Blanco:2018esa} that 
limits on DM decay from the isotropic gamma-ray background for different annihilation channels 
tend to lie within a factor of about 2 of each other.  As noted above, DM annihilation in 
microhalos closely resembles DM decay, so the same would apply to limits on annihilation 
following an EMDE.

\end{itemize}

There are also gravitational search prospects for microhalos.  One should keep in 
mind here that microhalos are not as dense as other compact objects are: they're way more 
diffuse/puffy.  Imagine an Earth-mass object with its mass smeared out over a volume whose 
radius is comparable to that of the entire solar system and you'll have a fairly accurate 
picture of what microhalos are like.  One possible way of detecting these objects is through
the pulsar-timing anomalies they can induce when they pass near pulsars and perturb their 
motion.  These anomalies can potentially be observed at 
PTAs~\cite{Ramani:2020hdo,Lee:2020wfn,Blinov:2021axd,Delos:2021rqs}.  
People do this with PBHs~\cite{Dror:2019twh}, but this is easier to do with PBHs than it is 
to do with microhalos because PBHs are genuinely compact.  We'd need a population of 
$\mathcal{O}(10^3)$ millisecond pulsars in order to see a signal, and we currently have a 
catalogue of about $\mathcal{O}(70)$.  We'd also need way more timing precision than we 
currently have.  Cluster caustic microlensing~\cite{Diego:2017drh,Oguri:2017ock} is another 
technique that can be used for detecting microhalos~\cite{Blinov:2021axd}.  The idea here is 
to look at individual stars lensed by galaxy clusters.  Stars transiting behind caustics in 
the potential of these clusters get significantly magnified.  If a compact object --- in this 
case, a microhalo --- within the cluster causes a microlensing event, the magnification varies 
over time.  This technique can also be used with axion mini-clusters~\cite{Dai:2019lud}.  
There's also a very futuristic idea that one might be able to use interferometry with fast 
radio bursts to detect microhalos if one had an interferometer roughly the size of Saturn's 
orbit~\cite{Xiao:2024qay}.

EMDEs can also give rise to GW signals.  As was discussed earlier in this workshop, 
density perturbations source tensor perturbations at second order.  If the decay rate of the 
metastable field that's dominating the energy density of the universe during the EMDE is 
rapid compared to $H$, this can lead to enhancements in the GW spectrum.  
In the non-linear regime, halo collapse during an EMDE can also source GWs.  The halos that
are the last to form yield the largest contribution.  This is both because these halos are 
the largest in size and because the energy density $\rho_{\rm GW} $ of the GWs they produce 
scales with $a$ like radiation ($\rho_{\rm GW} \propto a^{-4}$), which means that the 
corresponding abundance will be suppressed more during a MD era (the EMDE) than during a 
RD one.  Such GW signals could possibly be detectable by far-future 
experiments~\cite{Lozanov:2023knf,Fernandez:2023ddy}.


\section{Block~V: Scalar Fields and Non-Standard Expansion Histories II (Specific Models)\label{sec:block5}}


\begin{center}
{\bf Discussion Leaders}\/: \\ Akshay Ghalsasi and Tristan L.~Smith 
\end{center}

Given some results that have appeared in the literature within the last month or
so~\cite{Freedman:2024eph,Riess:2024vfa}, the question that I think I need to address 
before we deal with anything else is this one: is there still a Hubble tension?  The short 
answer is yes.  It hasn't gone away.  Moreover, the phrase ``Hubble tension'' is a misnomer: 
what we're really contending with is a bundle of interrelated tensions between different
measurements~\cite{Aylor:2018drw,Knox:2019rjx,Bernal:2021yli,Poulin:2024ken,Pedrotti:2024kpn}.  
A better name for this issue would be the ``cosmic calibration tension.''

The angular size/scale/structure of the CMB must remain constant no matter how you
modify cosmology.  In particular, we need $\theta_s$ (the angular size of the sound 
horizon) to be the same in any cosmology.  At recombination, $\theta_s$ can in general 
be approximated by
\begin{equation}
  \theta_s ~=~ \frac{r_s(z_{\rm rec})}{D_A(z_{\rm rec})}  
    ~\sim~ \frac{H_0}{H(z_{\rm rec})} \frac{c_s(z_{\rm rec})}{F(\Omega_m)}~,
\end{equation}
where $r_s(z)$ is the sound horizon, $D_A(z)$ is the angular-diameter distance, $H_0$ is the 
present-day Hubble parameter, and $F(\Omega_M)$ is a factor which depends on the total 
present-day matter density $\Omega_M$, but is to a good approximation redshift-independent 
at high $z$.  On the other hand, the Silk-dampling scale is given by 
\begin{equation}
  \theta_d ~\sim~ \frac{H_0}{\sqrt{\dot \tau(z_{\rm rec}) H(z_{\rm rec})}}~,
\end{equation}
where ${\dot \tau}(z_{\rm rec}) \equiv n_e(z_{\rm rec})x_e(z_{\rm rec})\sigma_T /(1 + z_{\rm rec})$ is 
given in terms of the electron number density $n_e(z_{\rm rec})$ and ionization fraction 
$x_e(z_{\rm rec})$ at recombination and in terms of the Thomson cross-section $\sigma_T$.
This means that if $H(z_{\rm rec})$ increases while keeping $\theta_s$ fixed, then the 
angular Silk-damping scale must also increase.  This, in turn, means that $n_s$ generically 
also increases~\cite{Poulin:2023lkg}.

Now let's take a look at the effect that early dark energy (EDE) has on the
cosmic-calibration tension.  For concreteness, let's consider an {\it ad hoc}\/  
model where the scalar potential is
\begin{equation}
  V(\phi) ~\propto~ m^2f^2[1 - \cos(\phi/f)]^3~. 
\end{equation}

\begin{itemize}

\item Question: Is this motivated?  

\item Answer: This form for $V(\phi)$ is fairly {\it ad hoc}\/.~  Really, the motivation 
is bottom-up and not top-down.  Potentials of the form $V(\phi) \propto m^2f^2[1 - \cos(\phi/f)]$
(without the cube) or of the form $V(\phi) \propto |\phi|^n$ don't work as well. 

\item Comment: A potential like this doesn't really arise in string theory.  
It would only come from instanton effects at higher order, and the contributions to the 
scalar potential from these effects are quite suppressed.

\item Question: Is there an axion quality problem in this
case, given this form of the potential?  Wouldn't it be hard to protect the shape of
$V(\phi)$ from being perturbed?

\item Answer: Yes, in general that is true.  As was commented on before, this form for $V(\phi)$ 
is {\it ad hoc}\/.~  However, that being said, there are some examples in which potentials 
of this form are produced within a string-theory context 
(see, \eg, Refs.~\cite{McDonough:2022pku,Cicoli:2023qri}).

\end{itemize}

We can also consider another class of EDE models in which the universe is 
initially trapped in a metastable vacuum and then undergoes a first-order phase transition 
to the ground state.  One simple example is the so-called ``new early dark energy'' (NEDE)
model~\cite{Niedermann:2019olb,Niedermann:2020dwg}, which involves two scalar fields: a 
dark-energy field $\psi$ and a ``trigger'' field $\phi$ with masses $M$ and $m$, respectively.  
In order to address the cosmic calibration tension, we need $M\sim 1$~eV and 
$m \sim H_{\rm eq} \sim 10^{-27}$~eV.  The EDE and NEDE models yield similar predictions 
for $H_0$ and similar values for most cosmological parameters, but they yield 
different values for $S_8$~\cite{Poulin:2021bjr}.

\begin{itemize}

\item Question: What's the physical motivation for the mass/energy scales of the 
dimensionful parameters that characterize these scalar models?

\item Answer: One interesting possibility is that they could be related to neutrino masses.  
Neutrino decoupling happens at similar energy scales, so there would be a natural connection 
to make here.

\item Comment: People have looked at this, actually~\cite{Escudero:2019gvw,
Escudero:2021rfi,Arias-Aragon:2020qip,Fernandez-Martinez:2021ypo}.  Indeed, most proposals 
for addressing the Hubble tension have a coincidence problem in the sense that they require 
that the standard cosmology be modified when the temperature of the radiation bath 
is right around the recombination scale ($T \sim 0.1$ -- $1$~eV).  There is no 
motivation in the SM (or even any anthropic motivation) for any cosmological 
dynamics to be triggered at this seemingly arbitrary temperature scale.  The only 
interesting thing that happens around this scale is that neutrinos become non-relativistic,
and so people have tried to relate solutions to the Hubble tension along these lines to BSM 
modifications of neutrino cosmology.  One concrete way of establishing a connection of this sort 
in the context of EDE is to posit that dark energy is sourced by neutrinos 
with time-varying masses which couple to a scalar field~\cite{Kaplan:2004dq,Fardon:2003eh}.  
The dark energy disappears after the neutrinos become non-relativistic, and indeed this indeed 
happens at the appropriate scale to address the Hubble tension~\cite{Gogoi:2020qif}.

\end{itemize}

For the purpose of disentangling perturbation effects from background 
effects, it's also useful to look at fluid models (rather than particle-physics-motivated
models) for dark energy~\cite{Lin:2019qug,Poulin:2023lkg}.  One of the primary lessons 
from these kinds of studies is that perturbations really are important.

If I had to distill one take-away message out of all of this, it would be as follows. 
Beware analyses that don't take into account all three of the following: supernovae, BAO, 
and Planck data.

\begin{itemize}

\item Question: Why not leave out supernovae and leave room to modify the 
input from SH0ES, \etc, in light of possible systematic uncertainties?  

\item Answer: That's a good point.  One could do that in principle, and I certainly think 
there's motivation for doing that.

\end{itemize}

Now let's switch gears a bit and talk about another possibility which can arise
in theories with additional light scalars, which is a kination epoch.  I'll focus on a 
particular kination scenario here, which is axion kination~\cite{Co:2019wyp}.  The idea is 
that when an axion is rolling along the flat direction at the bottom of the canonical 
Mexican-hat potential, its energy density scales with $a$ like kination.  Higher-dimensional
operators explicitly break the Peccei-Quinn symmetry and disturb the flatness of the 
potential.  

\begin{itemize}

\item Question: This setup is intimately tied to Affleck-Dine 
baryogenesis~\cite{Affleck:1984fy} and the related work of Dine, Randall, and 
Thomas~\cite{Dine:1995kz} back in the mid-1990s.  Why are people revisiting this now?

\item Answer: It's actually more closely related to the mechanism known as spontaneous
baryogenesis~\cite{Cohen:1987vi,Cohen:1988kt}, wherein a chemical potential is generated 
for baryon number.  Indeed, baryogenesis from the rapid rotation of axions is a 
realization of spontaneous baryogenesis.  Spontaneous baryogenesis is distinct from 
Affleck-Dine baryogenesis, wherein baryogenesis arises due to the decay of the the 
($CP$-violating) complex-scalar condensate.  The renewed interest in this baryogenesis 
mechanism is probably due to the increase in interest within the community in the physics 
of axion-like particles over the last half decade or so.  All that models along these
lines need is a large initial displacement and small explicit symmetry breaking at large 
field values to provide the initial velocity kick. 

\item Question: Would Hubble friction damp the rotational motion of the 
axion in field space?  

\item Answer: Yes, it would.  It does.  This is part of the model.  

\end{itemize}

Initially, the axion rolls down the potential {\it while}\/ it rotates.  It has both 
potential and kinetic energy, and the energy density of the field scales like that of
matter.  However, later, when the field is just rotating around the trough of the potential,
it has no potential energy, just kinetic energy, so it acts like kination.  Thus, the cosmological
picture here is that we have a MD epoch followed by a kination epoch.  

This phenomenon can in principle be incorporated into the cosmological timeline 
at a lot of different points during the history of the universe.  In particular, it can
occur either before or after BBN.

\begin{itemize}

\item Comment: The tough thing to deal with in modified cosmological 
histories after BBN is typically the ``graceful exit'' that's needed when the epoch ends. 
This often involves a substantial entropy dump.  This is the case, for example, with an EMDE.  
This model is one which works, but this consideration imposes very stringent constraints.  

\item Question: Should we have concerns regarding the fragmentation of the
field condensate here?  In models where one tries to do similar things with the inflaton, 
there {\it is}\/ fragmentation, and this changes the story quite a bit.

\item Answer: Yes the axion condensate is expected to fragment due to parametric 
resonances, which lead to instabilities.  Many of the implications were worked out in 
detail in Refs.~\cite{Fonseca:2019ypl,Eroncel:2022vjg}.  An additional contribution to 
the GW background can arise as a consequence of fragmentation.

\item Question: Are there questions about how sphalerons enter into 
this picture?  Do we know how to calculate the sphaleron rate?  

\item Answer: At this point, yes, we do have a sense of how sphalerons 
can matter.  Sphalerons allow us to re-interpret the relevant term as a chemical potential, 
as in spontaneous baryogenesis~\cite{Cohen:1991iu,Giudice:1993bb}.  An example of a modern 
calculation in this type of axion model including sphaleron effects can be found in 
Ref.~\cite{Domcke:2020kcp}.  However, it was argued in Ref.~\cite{Dolgov:1996qq} that   
this interpretation was invalid, at least in the spontaenous-baryogenesis scenario. 
I don't have anything to say about the sphaleron-rate calculation, though.  

\item Question: How generic are the initial conditions that give rise to 
a setup like this?  Should we worry about kination epochs all over the place if there's a
string axiverse~\cite{Arvanitaki:2009fg}?  

\item Answer: All you really need is a large initial field value for this 
to happen.  That would suggest that indeed it's pretty generic.  

\item Comment: This picture might get complicated in scenarios in which 
there are multiple axions --- for example, in KK theories with bulk axions or in the 
axiverse --- because there can be mixing among axions.  That mixing would need to be
accounted for.  Sometimes this sort of axion mixing can lead to unexpectedly large 
resonance-type effects, particularly when phase transitions are
involved~\cite{Dienes:2015bka,Dienes:2016zfr}.

\item Comment: One might worry if the vacuum expectation values (VEVs) of the fields 
needed to be trans-Planckian in order for this to work, but they needn't be.  

\end{itemize}

There's an enhancement to the GW spectrum from high-scale kination.  As other people 
have discussed already during this workshop, the energy density associated with GWs with a 
given wavenumber $k$ scales with $a$ like radiation once those GWs enter the horizon.  
During a kination epoch, just like during an EMDE, the radiation energy density constitutes 
a much smaller fraction of the critical density than it does during a RD 
epoch, and the power in GWs in the corresponding frequency band in enhanced.

If there's a kination component contributing to the energy density of the universe after
BBN, it can help address the cosmic calibration tension.  The interesting thing here is that
late kination can reduce the $H_0$ tension without increasing the $S_8$ tension.


\section{Block~VI: How Inflation Ends: Non-Standard Possibilities\label{sec:block6}}


\begin{center}
{\bf Discussion Leaders}\/: \\ Mustafa Amin and John T.~Giblin, Jr.~ 
\end{center}

Dynamics during the cosmological epoch which follows inflation can have a significant 
impact on the subsequent cosmological history.  Observational signals that could potentially 
tell us something about these dynamics include GWs, the signals associated with relic populations 
of different particle species (for example, those that constitute the DM), 
and the properties of power spectra that show up in the CMB and large-scale structure.  
The effective equation-of-state parameter $w$ for the universe tells us how the universe 
expands at any given time.  Inflationary observables such as $r$ and $n_s$ are affected by 
the value that $w$ takes during the period following inflation, as is the differential primordial 
abundance $d\Omega_{\rm GW}(f)/df$ of GWs as a function of their frequency $f$. 

Data from Planck favors a flatter than quadratic inflaton potential 
$V(|\phi|)\propto |\phi|^{\alpha}$, with $\alpha<2$ at large field values~\cite{Planck:2018jri}.  
Since the whole potential can no longer be quadratic for all field values, it is natural to parametrize 
the inflaton potential near its minimum as $V(|\phi|) \propto |\phi|^{2n}$, which then flattens 
out to the aforementioned $V(|\phi|)\propto |\phi|^{\alpha}$ for $|\phi|>M$.  Here, $M$ is 
some scale (\ie, some field distance from the potential minimum) beyond which the power law changes 
from $2n\ge 2$ to $\alpha<1$.  We will treat $M$ and $n$ as free parameters in what follows.  
Typically, people consider values for these parameters within the regime in which $M\lesssim M_P$ 
and $n\ge 1$, though other values are not necessarily forbidden.  Although $\phi$ typically only 
has sub-Planckian excursions after the end of inflation, during inflation this field can evolve 
over super-Planckian ranges.

\begin{itemize}

\item Now, before I go any further, I'd like to pose a question to everyone.  
What do we think about trans-Planckian field VEVs?  Are we comfortable with them?  

Answer: Such VEVs are worrisome, at least from an EFT perspective.  From a string 
perspective, it's often also a problem.  However, in models like those motivated by axion 
monodromy, there are perturbative shift symmetries that protect the field-theory Lagrangian 
against corrections.  In such scenarios, having $|\phi| \sim \mathcal{O}(M_P)$ wouldn't be 
too much of an issue.  However, if $|\phi|$ is {\it parametrically}\/ larger than $M_P$, 
that probably still {\it would}\/ be an issue.

\end{itemize}

Okay, well, let's first consider what happens when $M \sim M_P$.  If we have a quadratic 
potential ($n=1$) at the minimum, after inflation ends $|\phi_{\rm end}|\lesssim 0.1M_P$, we 
do not have access to the non-gravitational self-interactions that appear at $|\phi|\sim M$. 
The field oscillates at a frequency corresponding the inflaton mass $m$, with a 
time-averaged equation-of-state parameter $\langle w \rangle = 0$. This coherently oscillating 
field acts like non-relativistic matter.  On long time scales (in comparison with the oscillation 
period $m^{-1}$), inhomogeneities grow due to gravitational interactions, with $\delta\propto a$. 
These inhomogeneities can eventually become nonlinear, similar to what happens during the 
usual MD epoch that occurs after MRE --- and they can also form gravitationally supported
solitons~\cite{Easther:2010mr,Niemeyer:2019gab}.

Now if $n=1$, and $M\ll M_P$, at the end of inflation $\phi$ has significant non-gravitational 
self-interactions (since $\phi_{\rm end}\gtrsim M$).  The oscillating field then has a Floquet 
instability at a characteristic wave number $k\lesssim m$.  The instability timescale is 
short compared to the Hubble time --- \ie, $t_{\rm inst.}/t_H\sim (M/M_P)^2\ll 1$ --- and thus 
the associated dynamical effects are fast compared to gravitational clustering.  
As a result, we get rapid fragmentation of the homogeneous 
inflaton field into solitons (``oscillons'')~\cite{Amin:2011hj}.  This initial rapid formation 
of oscillons isn't a result of gravitationally-induced clumping; it's a result of inflaton 
self-coupling-induced clumping.  However, in spite of there being large inhomogeneities present, 
you don't in general get huge gravitational potentials for these solitons ($|\Phi|/c^2\ll 1$), 
so you don't have them collapsing and turning into PBHs 
easily~\cite{Lozanov:2019ylm,Ballesteros:2024hhq}.  Nevertheless, as time 
goes on, the oscillons themselves do also clump gravitationally~\cite{Amin:2019ums}.  Overall, 
in spite of the rich dynamics involved, the temporally and spatially averaged equation of state 
still is $\langle w\rangle \approx 0$.

\begin{itemize}

\item Question: How likely is it that we'd actually get a parametric resonance in
a setup like this?  

\item Answer: It's probably more likely than one might na\"{i}vely think. The perturbations 
of the field around the oscillating background has periodic time-varying effective frequency 
(since the potential is not purely quadratic, it has ``wings'').  So parametric resonance is 
possible.  Now, for the resonance to be effective in generating large inhomogeities, the timescale 
for the growth of the resonant instability must be fast compared to the Hubble scale.  This happens 
whenever $M\lesssim 0.1 M_P$.  Separately, even purely gravitational clustering (which occurs when 
$M\sim M_P$) in the linear regime can be thought of as being due to a resonant growth.  It's just that
when expansion is taken into account, the growth in this case is power-law growth rather than 
exponential growth.

\item Question: Can these solitonic clumps attract each other?  Can they become
gravitationally bound into binaries and inspiral?  

\item Answer: Yes, they gravitationally clump together, forming binaries, \etc~\cite{Amin:2019ums}. 
However, they're field configurations with spatial extent, not point particles, so what happens when
they overlap depends on the particular field configurations in the clump.  In particular, the relative 
temporal phase difference of the field between clumps matters.  In some cases, the clumps will 
bounce off each other; in other cases, they'll merge, with a merger being the more likely outcome. 

\end{itemize}

So that's the situation for $n=1$.  Now let's generalize to a non-quadratic power law. 
One obtains different expansion histories depending on the value of $n$.  Indeed, for 
different $n$, you get different equations of state [for the coherent oscillations of 
the homogeneous background, $\langle w\rangle= (n-1)/(n+1)$]~\cite{Turner:1983he}. 
For $n\ne 1$, we also get fragmentation of the field regardless of whether $M\sim M_P$ or 
$M\ll M_P$ due to self-resonance.

\begin{itemize}

\item Now, since asserting that $n > 1$ means that the quadratic term in $V(\phi)$ is so 
small as to be negligible here, I have to pose another question.  From a model-building or 
field-theoretic perspective, is it reasonable to have this mass/quadratic term be small 
and just have a $V(\phi) \sim \lambda |\phi|^4$ potential?  

\item Answer: Sure!  After all, the Higgs potential has that property, and the Higgs 
boson exists.

\end{itemize}

Okay, in that case, let's take a look at what emerges for for $n > 1$.  For all $n>1$ 
(not just particular values, or integers), fragmentation of the homogeneous field is 
inevitable due to self-resonance.  The instability always becomes faster than the Hubble 
scale eventually, even if it is not the case initially (the time when this happens is 
proportional to $n$ for $n\ne 1,2$).  After fragmentation, since the field quanta are 
massless at the bottom of the potential when $V''(\phi\rightarrow 0)\rightarrow 0$, we 
end up with a relativistic gas of massless particles with an equation of state $w = 1/3$. 
So for $n > 1$, we eventually end up with a ``radiation'' dominated universe even without 
the inflaton coupling to other fields~\cite{Lozanov:2017hjm}.  Adding additional light 
fields only speeds up the process~\cite{Antusch:2020iyq}. 

\begin{itemize}

\item Comment: Won't there be corrections at loop order to the monomial 
form of $V(\phi)$ that you're asserting?  

\item Answer: Yes, there will be, but people have considered these quantum corrections, 
and it turns out that the story doesn't change a lot at early times when you take them into 
account.  For example, you can get a quadratic term at loop level, but if the coefficient 
$m_\phi^2$ for this term is smaller than the second derivative $V''(\phi)$ of the tree-level 
potential with respect to $\phi$ (for $\phi\ne 0$), then it does not matter as far as the 
self-resonance is concerned.  Eventually when the field value becomes sufficiently small, 
or when the momenta of the quanta produced in this way redshift enough, the small mass will become 
important. Thereafter the universe would again move towards matter domination.  In terms of 
decays to other fields, there can be relevant effects~\cite{Garcia:2023eol}.

\item Question: What's the phase-space distribution of the fragmented field 
in the $n=2$ case?  Is this distribution just peaked around the resonance?  

\item Answer: It starts our with a sharp resonant peak.  Because of the interaction term in 
the potential --- this is, after all, a potential dominated by a quartic term near the 
minimum --- the field quanta interact and scatter with each other, so their momenta get 
redistributed, which then creates harmonics via re-scattering and broadens the phase space 
distribution via mode coupling.  The final distribution is not an incredibly sharply peaked 
distribution, and it's not a thermal distribution either~\cite{Khlebnikov:1996mc,Lozanov:2017hjm}. 

\end{itemize}

This brings us to the topic of preheating --- the rapid transfer of energy from from the 
inflaton to other fields.  Historically, nonlinear physics has been invoked to quickly and 
efficiently move the universe toward a RD ($w=1/3$) state.  Now, what we've 
learned is that couplings between the inflaton and other fields gives rise to a rich 
phenomenology and it's not so simple to turn off all the processes that lead to rapid preheating.

\begin{itemize}

\item Question: How does this translate into alternatives to inflation?  
The energy scale of inflation that people are interested in seems to be getting lower and lower
over time.  How to we get to a RD universe after inflation ends as quickly as possible?  

\item Question: To follow up on that question, is it actually hard to get the
universe into a RD epoch after inflation?  

\item Answer: I'll address these two questions as follows.  Preheating actually seems to happen all 
the time --- or almost all the time --- and it works very well in the sense that it can 
give us a RD universe within a acceptable time frame (especially relevant for 
very-low-energy-scale inflation).  By contrast, it may be harder to obtain a RD universe within 
an acceptable time frame via perturbative reheating.

\end{itemize}

One example of very efficient preheating occurs in what's called gauge preheating, wherein 
$\phi$ couples to gauge-invariant combinations of gauge-field operators via dilatonic 
interactions of the form $W(\phi)F^{\mu\nu}F_{\mu\nu}$ that modify the kinetic terms of the 
gauge fields, or via axial interactions of the form 
$X(\phi)\tilde{F}^{\mu\nu}F_{\mu\nu}$~\cite{Adshead:2015pva}, where $W(\phi)$ and $X(\phi)$ are
general functions of $\phi$.  In models of this sort, one can get density contrasts of order 
$\delta_k \sim \mathcal{O}(1)$ for certain $k$--often at wavelengths near the horizon--right 
after production~\cite{Adshead:2023mvt}.  However, the fields that are being produced are 
massless, so the clumps won't grow unless they do in fact collapse and form PBHs, and the 
fragmented clumps aren't dense enough to do that.  Thus, we {\it still}\/ get a RD universe 
at the end.  

We can also consider particular forms for the potential which are motivated by explicit 
top-down constructions --- for example $\alpha$-attractors~\cite{Kallosh:2013hoa}.  
In this case, we have a particular functional form for $V(\phi)$ and we get a derivative 
coupling of the inflaton to an axion field.  These $\alpha$-attractor potentials generally 
{\it still}\/ give us efficient preheating.  In general, one also finds that a lot of 
GW production occurs for these $\alpha$-attractor potentials.  The results are only weakly 
model-dependent.

While we've been focusing on the inflaton field here, its important to emphasize that a
scalar field that gives rise to EDE behaves in the same way.  The coherent, homogeneous 
background can fragment, and GWs can be produced as a result.

\begin{itemize}

\item Question: So why doesn't the QCD axion do this?  Why doesn't its 
condensate fragment?  

\item Comment: Doesn't it?  Isn't this fragmentation related to how axion stars 
arise?

\item Answer: Assuming global misalignment, whether or not fragmentation 
occurs for an axion-like field with a cosine potential during radiation domination (without 
coupling to other fields) depends on the ratio $\phi/f$, where $f$ is the axion-decay constant.  
If this ratio gets as large as $\phi/f \sim \pi$, then you could potentially get fragmentation 
of the condensate at early times purely due to self-interactions~\cite{Arvanitaki:2019rax}.  
However, if the field is rolling during an RD era and the condensate isn't dominating the universe, 
the dynamics is modified and its harder to achieve this.  With local misalignment, there is also 
string formation, which makes the dynamics richer and can lead to significant axion-star production. 

\item Question: Can you get warm inflation from these kinds of couplings?  

\item Answer: Yes, even if $\phi$ is just rolling, and there is a coupling to gauge fields 
especially in the presence of a $X(\phi)\tilde{F}^{\mu\nu}F_{\mu \nu}$ coupling, there's 
tachyonic instability in the gauge fields, so you can get particle production during inflation 
before oscillation begins.  You might obtain extra damping in this way and slow the rolling of 
the inflaton down to the extent that you can get warm inflation, so yes, it may be possible.  

\item Comment: Arranging that might require a bit of 
fine-tuning~\cite{Yokoyama:1998ju}.  

\item Comment: What if you couple the inflaton to fermions instead of bosons?  
Won't a lot of things change?  To what extent will they change?  For example, what if you 
coupled the scalar to right-handed neutrinos?

\item Answer: When fermions are produced from a homogeneous inflaton field, the produced 
fermionic field cannot have large occupation numbers. As a result there need not be any 
significant back-reaction on the inflaton.  However, there can be some differences from 
perturbative calculations (see, \eg, Refs.~\cite{Greene:2000ew,Adshead:2015kza}).

\end{itemize}

In multi-field models, many of the quantitative outcomes of the post-inflationary dynamics 
are sensitive to the sizes of Lagrangian couplings, as well as initial 
conditions~\cite{vandeVis:2020qcp}.  Nevertheless, there can be some surprising universality 
in the dynamics if the number of decay products is large or the field-space metric is 
sufficiently complicated~\cite{Amin:2015ftc}.


\section{Block~VII: Connections to Formal Theory\label{sec:block7}}


\begin{center}
{\bf Discussion Leaders}\/: \\ James Halverson and Gary Shiu 
\end{center}

Cosmological acceleration, as we've discussed already many times during this workshop, 
is often driven by scalar fields (\eg, moduli).  String theory involves a lot of scalars 
which describe the geometry and properties of the compactification.  These scalars need 
to be stabilized.  This issue was originally known more generally as the Polonyi problem 
(see Ref.~\cite{Kane:2015jia} and references therein).  If these fields are not stabilized 
or given masses sufficiently large that they decay prior to BBN, they can lead to 
time-varying coupling constants and violations of fifth-force constraints.  These scalars 
could also affect BBN observations if they decay after BBN.  On the other hand, depending 
on their masses, these string-theory moduli/axions can also lead to EMDEs or to cosmological 
stasis, and can have a number of observational consequences.  They can also, somewhat 
paradoxically, address the {\it cosmological moduli problem}\/, since that the entropy 
injection from their decays at the end of an EMDE can dilute the abundances associated with 
pre-existing populations of {\it other}\/ moduli~\cite{Kane:2015jia}.  For recent discussions 
of the challenges and opportunities afforded by scalars in string theory, see, \eg,
Refs.~\cite{Flauger:2022hie,Cvetic:2022fnv,Marchesano:2024gul}.

One may ask: if string theory introduces all these extra ingredients, why bother? 
A main advantage of this top-down approach is that it narrows the infinite possibilities of 
quantum field theory (in the presence of gravity) down to a more restrictive set of 
possibilities --- through the imposition of consistency conditions.  Some of these 
consistency criteria are familiar from quantum field theory --- \eg, causality and 
unitarity bounds on the Wilson coefficients~\cite{Adams:2006sv} though these bounds 
are modified by gravity~\cite{Hamada:2018dde}.  Other, more subtle quantum-gravity 
criteria which constrain the range of field-theoretic models can be inferred from 
BH considerations.

In string theory, coupling constants are determined by the VEVs of scalar fields.  
A vacuum state exists only if terms involving different powers of the fields compete.  
Realizing a de Sitter vacuum requires at least three competing terms.  However, if 
competing terms of different order in the coupling expansion compete, why don't even 
higher order terms matter?  This is the gist of the Dine-Seiberg problem~\cite{Dine:1985he}.
Symmetries can forbid a few operators: for example, they can be used to ensure the 
smallness of the slow-roll parameters (which are sensitive to dimension-6 Planck-suppressed 
operators) associated with cosmic inflation.  The challenge presented by the Dine-Seiberg 
problem is how to have control, in principle, over all terms in the asymptotic expansion 
of the couplings.

Observations suggest we have two different periods of accelerated expansion.  There's 
cosmic inflation, which occurs early in the history of the universe, and there's the present 
epoch during which dark energy drives the expansion.  Of particular relevance for observation 
is the Lyth bound~\cite{Lyth:1996im}, which basically implies that models of single-field 
inflation that give rise to detectable GW signals require the inflaton potential $V(\phi)$  
to be nearly flat over a super-Planckian range of field values.

The current period of dark-energy-fueled acceleration can be arranged by virtue of a 
de Sitter minimum, a de Sitter maximum, or a runaway potential with the Hubble slow-roll 
parameter $\epsilon \equiv -\dot{H}/H^2 < 1$.  In contrast with inflation, which needs to 
last 60 or more $e$-folds, there is no real limit --- beyond an $e$-fold or two --- on 
the number of $e$-folds that the current dark-energy period needs to last.  Dark energy 
sourced by a field rolling down its potential is known as quintessence.  There are 
fifth-force constraints on quintessence~\cite{Jain:2010ka}, but they are model-dependent.  
In particular, these constraints depend on the couplings between the quintessence field 
and the SM particles.  Axions with only derivative couplings to the SM particles can evade 
fifth-force constraints. Thus they are often employed in quintessence model building.

\begin{itemize}

\item Comment: A cosmological constant is still viable, and measurements of the 
present-day equation-of-state parameter for the universe~\cite{DESI:2024mwx} can potentially 
be important for distinguishing this possibility from quintessence, although modifications 
of gravity are still possible and their observational proof must be carefully 
analyzed \cite{Wen:2021bsc}. 

\end{itemize}

One possible way of realizing quintessence in string theory is to posit that our universe
inhabits certain asymptotic regions of the string landscape.  Within these asymptotic regions,
towers of states become light~\cite{Ooguri:2006in}.  These towers comprise large numbers of 
individual degrees of freedom.  Through use of the covariant entropy bound, it has been shown 
that the potential has a universal exponential fall-off in any asymptotic region of the 
landscape~\cite{Ooguri:2018wrx}.

\begin{itemize}

\item Question: Does gravity get weak in those asymptotic regions of the landscape?  

\item Answer: Yes, the gravitational coupling gets weaker and weaker as we approach the 
asymptotic limit.  However, this asymptotic behavior is more general, since it applies to 
other couplings controlled by the VEVs of other moduli which may not have anything to do 
with the gravitational coupling.  The various small couplings observed in nature may point 
us toward such asymptotic regions.

\item Question: How robust are the correlations between small couplings, 
the presence of light towers of states, and stabilizing or rolling solutions for $V(\phi)$ 
that can give you late dark energy within these asymptotic regions of the landscape?  

\item Answer: The correlations between small couplings and the presence of light towers 
of states and an exponential falloff of the potential in the asymptotic regions of the 
landscape are robust.  The question is whether the potential has the right properties 
to give rise to acceleration.  As to whether the assumption of being in the asymptotic 
region is justified, one may draw an analogy with $1/N$ expansions in large-$N$ gauge 
theories.

\item Question: What you're describing is basically the distance conjecture.  
How well do we actually know that these expansions are reliable?  

\item Answer: If $\phi \gtrsim M_P$, then one has to worry about the reliability of the 
expansions.  The distance conjecture formalizes this worry by giving the physical reason 
for the breakdown of the EFT, as there are new degrees of freedom (towers of light states) 
not present in the original description.  Returning to the question of the accelerated
expansion that the universe is currently experiencing, the question is really whether we can 
bound the slow-roll parameter $\epsilon$ (defined as above), and doing so requires knowledge 
of the dynamics of (multiple) scalars.  Without knowing what combination of fields is going 
to give us the scalar degree of freedom that gives rise to dark energy dynamically, how 
do we bound $\epsilon$?

\end{itemize}

One can also consider multi-field quintessence models.  In such models, the potential 
$V(\phi_i)$ for the fields $\phi_i$ is a sum of exponential factors after one canonically 
normalizes the fields.  What appears in the exponents of these different factors depends on 
the relevant microphysics.  In models of this sort, one can bound the rate of time-variation 
of $H$ at late times \cite{Shiu:2023nph,Shiu:2023fhb,Shiu:2024sbe}.  Scaling solutions to the 
coupled equations of motion for $H$ and the $\phi_i$ --- \ie, solutions in which the scale 
factor $a$ grows with time as a power law --- turn out to be late-time attractors in these 
models. 

Now let's turn to the question of UV predictions in general.  Though not all field theories 
have known Lagrangian descriptions, we will restrict ourselves to the case of theories that do 
for simplicity.  With this setup, we make predictions in a UV theory by picking a vacuum, 
computing the associated Lagrangian, sometimes computing some intermediate data, and finally, 
from that information, we compute observables.  This is the ``forward model'' and it relies 
crucially on the vacuum selected, which requires a measure.  Conversely, given observables, 
one may ask how they constrain (via inference) the intermediate observables, Lagrangians, or 
UV theory, moving increasingly further back down the pipeline. Given the size of the string 
landscape (or really any cosmological theory with many vacua), this is a very hard inference 
problem.

Predictions depend crucially on picking a vacuum.  In the absence of a fundamental solution 
to the measure problem, one might instead simply ask what priors on Lagrangians arise 
naturally in string theory and how might they differ from the priors considered by bottom-up 
model-builders.  One important consideration here is that we expect many moduli,
axions~\cite{Arvanitaki:2009fg}, and gauge sectors (see, \eg, 
Refs.~\cite{Taylor:2015xtz, Halverson:2017ffz}), which can signficantly alter cosmology 
or particle physics --- \eg, due to the presence of dark
glueballs~\cite{Halverson:2016nfq,Acharya:2017szw}, 
ALP-photon couplings~\cite{Halverson:2019cmy}, or cosmological stasis~\cite{Halverson:2024oir}.  
We also expect KK modes, and in fact we are finally able to compute them~\cite{Ashmore:2021qdf} 
in smooth Calabi-Yau compactifications due to numerical metrics obtained with machine learning.  
See the TASI lectures~\cite{Halverson:2018vbo} on remnants from the string landscape for a 
thorough discussion of potentially observable remnants of string theory.

\begin{itemize}

\item Question: How do these numerical estimates inform distance-conjecture 
results for KK towers and their properties?  

\item Answer: There are indeed things that we can learn from these estimates.
Calabi-Yaus were originally identified as interesting because they yield SUSY below the KK 
scale.  If you want weak-scale SUSY --- which you might not necessarily want, but let's suppose 
for a moment that you do want it --- you'd want to compactify your string theory on a Calabi-Yau.  
Using numerical methods to obtain moduli-dependent Calabi-Yau metrics, one can compute the rate 
at which the KK states become exponentially light (see \eg, Ref.~\cite{Ashmore:2021qdf}) and confirm 
that the exponent is $\mathcal{O}(1)$, as suggested by the distance conjecture. 

\item Question: From a traditional particle-physics perspective, there's an 
insane amount of extra stuff in these models.  They yield an enormous hidden-sector content 
that in general is going to overclose the universe.  Why should we care about these models?  
 
\item Answer: I agree! Though moduli, axions, dark gauge sectors, and KK modes 
can give rise to new possible cosmological epochs, they can also place significant constraints 
on the space of string models.  For example, reheating all of the enormous number of gauge 
sectors in a typical F-theory compactification can give rise to too much dark 
matter~\cite{Halverson:2016nfq,Halverson:2019kna}.  One must carefully analyze both the 
observational opportunities and existing experimental constraints.

\item Comment: The string community also has its own prejudices about how 
to extract low-energy phenomenology from string theory.  Sometimes a lot of physics is hidden 
within the infinite towers of states that string theory provides, and not merely within the 
massless states. There is also a lot of focus on Calabi-Yaus, but what else is there beyond 
that?  To what extent must supersymmetry be the guiding force for model-building at the 
string scale?   

\end{itemize}

One can also perform data-set studies of the string axiverse in low-energy effective theories 
that emerge from these Calabi-Yau constructions~\cite{Demirtas:2020dbm,Gendler:2023kjt}.  
The results of the study suggest that both the masses $m_i$ and decay constants $f_i$
of the axions in these theories typically span broad ranges of values.  A more long-term goal 
in this direction is to compute the joint probability distributions for the $m_i$ and $f_i$ 
that emerge from individual such constructions, as this data and associated correlations 
are essential for understanding predictions in the axiverse.

\begin{itemize}

\item Comment: Another issue that's important to discuss is inference.  How do we infer 
information about the underlying theory from observables?  In other words, how do 
we --- and to what extant {\it can}\/ we --- address the ``inverse problem'' which which the
landscape presents us?

\item Question: The SM has 19 or so free parameters.  What kinds of 
measurements would be the most valuable in terms of pinning down the fundamental theory?  
Discovering an axion?  Measuring a coupling better?  

\item Answer: Inference problems on theory space are very hard in particle 
physics and cosmology, even at the level of the space of possible Lagrangians, long before 
we start worrying about the UV theory.  One has to work backward through the pipeline I 
outlined. Jonathan Heckman thought about the aspects (and difficulties) of this from a 
Bayesian-inference perspective~\cite{Heckman:2013kza}. To respond to the question that was
just asked, since (in my opinion) the prior from string theory involves many degrees of 
freedom, we should look for places where many degrees of freedom can lead to qualitatively 
different effects --- \eg, cosmological stasis.  That's why I'm excited about this meeting!

\item Comment: This is one of the biggest issues with the landscape.  
It's not clear that progress can be made on this inverse problem.  It's not really 
feasible to pin down the vacuum that was chosen by our universe.  For example, the 
landscape is so large that even if you stipulate a number of conditions for a viable 
string model, there are likely to be many models satisfying those conditions but yet 
differing in other important aspects.  This process may not converge, or converge 
in a reasonable way.  

\item Question: Aren't there some serious issues with sample bias in 
these kinds of landscape studies?  There are probably large classes of vacua that we're 
not even considering.  

\item Answer: To address the comment about how one might hope to address the inverse 
problem first, it would be better if we had a solution to the measure problem.  Regardless of that 
very hard problem, inference in string theory and on the space of possible Lagrangians is generally 
hard, because many models can give rise to the same types of observables (\cf\ the origin of simplified 
models).  To respond to the question that was just asked, in the absence of a non-perturbative 
definition of string theory that provides us with a knowledge of all vacua or sense of how to sample 
with the correct measure, we will always have sample bias.  In that case, the best you can do is explore 
as broadly as possible.  That's what's been happening for a number of years (\eg, in studies of large 
concrete ensembles in F-theory~\cite{Taylor:2015xtz, Halverson:2017ffz, Taylor:2017yqr} and weakly 
coupled Type IIB string theories~\cite{Demirtas:2018akl}).  The picture that is emerging is one of 
complexity: when we study compactifications as broadly as we know how, we typically see many moduli, 
axions, and gauge sectors.  

\item Comment: This is why the swampland program~\cite{Vafa:2005ui} is 
interesting.  One can in principle rule out classes of vacua --- \eg, those which yield irrational 
electrically charged particles~\cite{Shiu:2013wxa} or dark photons with masses above a certain 
scale~\cite{Reece:2018zvv}.

\end{itemize}


\section{Block~VIII: Cosmological Stasis\label{sec:block8}}


\begin{center}
{\bf Discussion Leader}\/: \\ Fei Huang 
\end{center}

The standard picture of the history of the universe is one in which the universe passes 
through a series of extended epochs, with the overall energy density during each epoch 
dominated by a single cosmological component (such as matter, radiation, or vacuum energy) with 
a particular value of $w$.  This is typically assumed to be the case simply 
because of the way the different energy densities associated with these components ultimately 
scale with the Friedmann-Robertson-Walker scale factor $a$.  Even if the epoch contains multiple 
kinds of energy components, very quickly one of them comes to dominate.  However, in the presence 
of the sorts of towers of states that arise within many BSM theories --- towers such as KK towers, 
which arise in theories of large extra dimensions and within which each state can be regarded as an 
individual particle species --- this expectation is often incorrect.   As a result, new kinds of 
cosmological histories are possible.

As an example, let's consider a scenario involving a tower of unstable states $\phi_\ell$
with masses $m_\ell$ which are non-relativistic and therefore behave like massive matter.  
Decays of these $\phi_\ell$ convert matter energy density to radiation energy density.  
The decay widths $\Gamma_\ell$ of the $\phi_\ell$ will scale across the tower in a particular 
way which is determined by the underlying theory, as will the primordial abundances 
$\Omega_\ell^{(0)}$ that these states have at some initial time $t^{(0)}$ before the most unstable of them
starts decaying appreciably.  The abundance associated with each $\phi_\ell$ decreases exponentially
as a consequence of particle decay, but because there are a large number of these fields and because each has 
a different decay width, the overall rate at which abundance is transferred from matter to radiation 
by the decays of the $\phi_\ell$ can end up scaling like $\sim 1/t$.  Since the Hubble parameter 
also scales with time like $H \sim 1/t$, it's possible that the effect of particle decay on the total
abundances $\Omega_M$ and $\Omega_\gamma$ of matter and radiation can actually compensate for the effect 
of cosmological redshifting, implying that $\Omega_M$ and $\Omega_\gamma$ can actually remain constant for 
an extended period.  

This phenomenon in which multiple different cosmological energy components have abundances which 
each remain constant despite cosmological expansion is called 
{\it cosmological stasis}\/~\cite{Dienes:2021woi,Dienes:2023ziv}.  Moreover, this stasis state in 
which the effects of the decays of the $\phi_\ell$ fields precisely cancel the effects of cosmological 
expansion as far as the abundances are concerned turns out to be a cosmological {\it attractor}\/.
Thus such a universe is pulled into a stasis epoch even if it did not start in one.  Thus far the stasis 
phenomenon has been observed in many different cosmological setups involving a variety of different 
models of BSM physics, a variety of different energy components, and a variety of different mechanisms 
(such as the particle-decay mechanism described above) whereby energy is transferred between the different 
energy components~\cite{Dienes:2021woi,Barrow:1991dn,Dienes:2022zgd,Dienes:2023ziv,Dienes:2024wnu,
Halverson:2024oir,Barber:2024vui}.

\begin{itemize}

\item Question: Is there an overall equation of state for the universe during 
stasis?  

\item Answer: Yes, it's an abundance-weighted average of the equation-of-state parameters 
$w_i$ of the individual cosmological components during stasis.  In the case we just described, these 
would be matter and radiation.  For example, if $\Omega_M = \Omega_\gamma = 1/2$ during such a 
matter/radiation stasis, we'd have $\overline{w} = 1/6$.

\end{itemize}

The manner in which the $\Gamma_\ell$ and $\Omega_\ell$ scale across the tower is 
important here.  Let's consider the case in which the $m_\ell$, $\Gamma_\ell$, and 
$\Omega_\ell^{(0)}$ scale with $\ell$ according to the scaling relations~\cite{Dienes:2021woi} 
\begin{eqnarray}
  m_\ell &\,=\,& m_0 + \ell^\delta \Delta m \nonumber  \\ 
  \Gamma_\ell &\,=\,& \Gamma_0 (m_\ell/m_0)^\gamma \nonumber \\
  \Omega_\ell &\,=\,& \Omega_0 (m_\ell/m_0)^\alpha~,  
\end{eqnarray}
where $\Delta m$ is a mass-splitting parameter and where $\delta$, $\gamma$, and $\alpha$ are
power-law indices.

\begin{itemize}

\item Question: Can you have a sizeable radiation bath, or is stasis 
contingent on the universe initially being MD at $t^{(0)}$?  Won't there be a sensitivity to
initial conditions?  

\item Answer: The normalization of the $\Omega_\ell^{(0)}$ will certainly depend 
on how much radiation is initially present at $t^{(0)}$, but the emergence of stasis isn't 
dependent on that.  The stasis solution is a dynamical attractor, and regardless of the 
initial values of the radiation and matter abundances, the system will flow toward stasis. 

\item Question: If stasis is a global attractor, how do you get out of it?

\item Comment:  Even though the stasis state is an attractor, there are many 
dynamical things happening ``under the hood'' in order to sustain it.  For example, 
in the case of the matter/radiation stasis with which we began this discussion, stasis 
requires the continual process of individual matter states decaying sequentially into
radiation.  In general, the heaviest states in the tower have the largest decay widths and 
decay first, then the slightly lighter states, and so forth sequentially down the tower.   
The stasis epoch thus ends naturally when you run out of states in the 
tower --- \ie, when the decay process reaches the bottom of the tower.  Indeed, as one 
reaches the bottom of the tower, numerous discretization effects appear which disturb 
the stasis and signal its impending end.

\item Question: How many states do you need to realize stasis?  If there are only two or 
three states in the tower that are populated, can you still get stasis?

\item Answer: We don't expect that stasis can be realized with only two or three states in 
this type of scenario, since stasis relies on sequential processes down the tower to balance 
against the cosmic expansion. In general, the number ${\cal N}_s$ of $e$-folds of stasis 
scales as the logarithm of the total mass ratio spanned between the top and bottom 
of the tower~\cite{Dienes:2021woi}.  Indeed, as shown in Ref.~\cite{Dienes:2021woi}, we 
generally have   ${\cal N}_s \sim \log(m_{N-1}/m_0)$.  For masses $m_\ell$ which grow as a 
power of $\ell$ this then produces ${\cal N}_s\sim \log \,N$, while for masses $m_\ell$ which 
grow exponentially with $\ell$ as in Ref.~\cite{Halverson:2024oir} this yields ${\cal N}_s\sim N$.

That said, there is an example of stasis~\cite{Barber:2024vui} in which no tower is needed 
at all.  This is a thermal stasis in which the energy transfer happens through particle 
{\it annihilation}\/ rather than decay.  Such a system nevertheless exhibits a fixed-point 
structure in which stasis emerges as an attractor, just as occurs in tower-based stases,  
with many $e$-folds of stasis being produced.

\item Question: Can one address the horizon problem using stasis?  

\item Answer: Yes, one can.  I haven't had a chance to talk about realizations of stasis 
involving other cosmological components beyond matter and radiation yet, but you can have a stasis 
involving vacuum energy and matter that gives rise to accelerated expansion. In such scenario, 
the inflationary epoch might actually be a stasis epoch.  We call this ``stasis inflation''.
You can easily get the appropriate number $\mathcal{N}\sim 60$ of $e$-folds of expansion 
necessary to solve the horizon problem in such a stasis.

\end{itemize}

On that note, it's worth pointing out that there can be other kinds of stasis that 
involve cosmological components other than matter and radiation~\cite{Dienes:2023ziv}.
For example, there can be a stasis between vacuum energy and matter.  In such a case, the 
canonical model would consist of a tower of scalar fields $\phi_\ell$ with different masses 
$m_\ell$.  Each field could transition from an overdamped state to an underdamped, oscillatory
state at a different time determined by the criterion $3H = 2m_\ell$.  The resulting epoch could then 
potentially be inflationary (with a stasis equation-of-state parameter $\overline{w}< -1/3$), and 
one could indeed solve the flatness 
and horizon problems this way~\cite{Dienes:2024wnu}.  A ``stasis inflation'' model along these lines 
would actually be interesting for many reasons.  Any stasis equation-of-state parameter 
$-1 < \overline{w} < -1/3$ can be realized and produce accelerated expansion, plus there's an automatic  
``graceful exit'' from inflation in scenarios like this because this inflationary stasis epoch 
ends when the underdamping transition reaches the bottom of the tower and all fields are oscillating. 
Moreover, a non-zero matter abundance is naturally sustained throughout the inflationary epoch. 
Finally, reheating could then potentially occur when our oscillating $\phi_\ell$ fields subsequently decay.  

\begin{itemize}

\item Question: In stasis inflation, what's determining the length of the stasis epoch and 
therefore the number of $e$-folds of inflation?  

\item Answer: In these kinds of stasis scenarios based on towers of BSM states, it's the 
range of $m_\ell$ in the mass spectrum of the scalars.  These are determining the range 
of times at which the fields transition from overdamped to underdamped oscillation.

\item Question: What if there are multiple decay pathways in a matter/radiation
stasis?  Is there an analogy to detailed balance and the way things work with stimulated 
{\it vs}\/.\ spontaneous emission?

\item Answer: As long as all of the possible decay pathways for the $\phi_\ell$ involve direct 
decays to final states comprising very light particles that behave like radiation across 
all relevant timescales, the existence of multiple such pathways won't have an impact on the 
stasis dynamics.  If the final states involve particles with non-negligible masses, 
then the situation becomes more complicated.  One has to consider the effect of cosmological 
redshifting on the equation of state for each massive particle species and therefore to 
consider not only how the overall abundances of these species evolve with time, but also how their
phase-space distributions evolve with time.  Effects such as these (but in a different context) 
were discussed in Ref.~\cite{Dienes:2020bmn}.~  The same would be true if there were non-trivial 
effects like stimulated emission which depend on occupation number.

\item Question: Is this stasis-inflation scenario related to 
$N$-flation~\cite{Liddle:1998jc,Dimopoulos:2005ac} in some way?  

\item Answer: They're closely related in many ways.  In stasis inflation, as in $N$-flation, 
you have a large number of different fields $\phi_\ell$ which have essentially separate potentials 
in the sense that $V(\phi_\ell)$ doesn't involve non-trivial couplings between different
$\phi_\ell$.   You might think the oscillatory $\phi_\ell$ fields are irrelevant after they 
start oscillating, since their energy densities fall rapidly with time during an epoch of 
accelerated expansion.  In stasis inflation, however, one has a {\it stasis}\/ between vacuum 
energy and matter (or some other energy component), which means that the effects of the matter are 
critical in sustaining the stasis.  While it is true that the individual energy density of each 
$\phi_\ell$ field will begin to inflate away once that field starts oscillating and thereby 
acting like matter (because the universe will necessarily have $w< -1/3$ during this epoch), 
there will always be another, slightly lighter field which which will then take its place.  
In this way stasis is maintained throughout the inflationary epoch until the bottom of the 
tower is reached.

\item Comment: $N$-flation is in the 
swampland~\cite{Brown:2015iha,Montero:2015ofa,Rudelius:2015xta}.
While a large number of fields may naively extend the effective field range, the latter is 
bounded by the Weak Gravity Conjecture~\cite{Arkani-Hamed:2006emk}, which requires extremal 
black holes to decay.  

\item Question: How do things typically decay in a KK tower in a top-down 
construction?  In, say, string constructions, do the $\phi_\ell$ typically decay to final 
states consisting entirely of particles outside the tower or to final states that include 
lighter $\phi_{\ell'}$ with $m_{\ell'} < m_\ell$ within the tower?  Is there a prediction?  

\item Answer: Both ``intra-ensemble'' (within the tower) and ``extra-ensemble'' 
(outside the tower) decay channels typically exist, and how the branching fractions work 
out is often highly model-dependent.  One must also be aware of situations in which there 
are symmetries which can restrict the intra-ensemble decays to only certain patterns.  For example, 
if our theory preserves translation invariance in the extra dimension, the corresponding KK states 
will respect conservation of KK number exactly (since this is nothing other than momentum 
conservation in the extra dimension).  If the masses of such states are also additive, this 
can render all such decays exactly marginal.   This is an example of the kinds of special 
model-specific situations that can arise.

\item Comment: In line with what was said a moment ago, it's worth 
emphasizing that it's also more challenging to analyze intra-ensemble decays numerically 
because you need to keep track of the evolution of the full phase-space distributions for the 
$\phi_\ell$, as the decay products can transition from the relativistic regime to the 
non-relativistic regime.  This increases the dimensionality of the problem.

\item Question: What about radiative corrections?  Won't those corrections 
affect the scaling relations?

\item Answer: Actually, we've taken a look at the robustness of stasis results 
against perturbations --- for example, perturbations to the mass matrix of the $\phi_\ell$ --- 
and you can quantify how robust things are.  As it turns out~\cite{Dienes:2023ziv}, 
they're pretty robust.  There's some jitter that arises in the otherwise constant abundances 
for the cosmological components that participate in stasis, but the overall qualitative results 
are unchanged.

\item Question: How does reheating work in stasis inflation?  There's no slow roll, but 
the end of inflation is basically the same as in single-field models: inflation still ends with 
an oscillating scalar --- or in this case a set of oscillating scalars.  It's like ``freeze-in'' 
reheating.  It's not exactly like what was being described earlier for single-field models because 
the initial conditions are different --- you aren't starting from a slow-roll conditions.

\item Answer: Yes, right.  Exactly.  

\item Comment: Anything that works for the usual reheating scenarios should 
also work for stasis inflation.

\item Comment: It is also possible that the stasis inflation is a triple stasis 
during which a non-vanishing radiation abundance is also maintained.  In this case, reheating 
happens naturally after all states in the tower decay away. 

\item Question: How does one think about reheating in string theory or theories 
with extra dimensions?  To what extent does one populate the towers?  There would presumably be 
some limit to the energy density, so you'll populate the states in the tower up to some value of
$m_\ell$ which is set by the Boltzmann-suppression factor.  Would you populate all sectors?  

\item Comment: Gravitational processes will populate everything.

\item Comment: We don't know if the mass spectrum for string axions is flat in 
$\log m_\ell$.

\item Comment: That's okay.  In general, what you really need for a long stasis is 
a hierarchy or large range of decay widths.

\item Question: Looking at the parametrized model, it seems that heavier states 
need to decay first.  Is this necessary?

\item Answer: It's typically like this if stasis is generated by particle decays.
However, there can be other ways to generate a matter/radiation stasis.  For example, you can 
get such a stasis from the evaporation of a population of PBHs with an extended mass
spectrum~\cite{Dienes:2022zgd}.  Hawking radiation in this case provides the ``pump'' 
which transfers energy density from matter to radiation.  In this scenario, the lighter PBHs 
are the ones that evaporate first, and during stasis the PBHs decay {\it up}\/ the tower.

\end{itemize}

Moreover, as I alluded to above, you can even get a matter/radiation stasis from a {\it single} thermal 
non-relativistic field~\cite{Barber:2024vui}.  No tower is needed, and the stasis arises not 
through particle decay but rather through annihilation to radiation.   The swept-volume rate 
for the annihilation process in such realizations of stasis takes the form 
$\sigma v \propto |\vecbf{p}_{\rm CM}|^q$, where $\vecbf{p}_{\rm CM}$ is the momentum of the 
incoming particles in the center-of-mass frame and where the power-law index $q$ is negative.

\begin{itemize}

\item Question: How do you get a negative value of $q$?  

\item Answer: There are a lot of ways to do this.  Sommerfeld enhancement~\cite{Sommerfeld:1931qaf} 
will do it, for example, but won't give you a sufficiently negative value of $q$.  It's not just 
that you need $q < 0$; you actually need $q$ to lie within a particular range of negative 
values~\cite{Barber:2024vui}.  However, there are other mechanisms that can give rise to negative 
$q$ within the desired range.

\item Comment: Don't you lose the beauty of the connection to KK constructions, 
extra dimensions and towers in realizations of stasis like that?  

\item Answer: The existence of such realizations of stasis is actually something that reflects 
a deeper mathematical structure of stasis.  

\item Question: Work has also been done --- by Jim Halverson and Sneh 
Pandya~\cite{Halverson:2024oir} --- on stasis in the axiverse  Can someone summarize 
the results?  

\item Answer: It's a very similar setup physically.  The emphasis in what we did 
was really on the numerical/machine-learning analysis that uncovered the particular realization
of stasis we found.  

\item Question: What if there's some numerical jitter due to a stochastic nature of
the mass spectrum or something --- what would happen?  Would you still get stasis?

\item Answer: As mentioned above, we've studied this, and 
stasis turns out empirically to be very robust.  Jim and Sneh's paper actually includes 
a number of figures which illustrate the effect of this jitter explicitly.  There's also 
a related phenomenon which we called ``oscillatory stasis''~\cite{Dienes:2023ziv}, which 
is stasis with ``wiggles'' in the 
abundances.  This can arise, for example, when the differences 
$\Delta \tau_\ell \equiv \tau_\ell - \tau_{\ell + 1}$ between the lifetimes $\tau_\ell$ of 
successive states in the tower are large in comparison with the corresponding lifetimes
themselves.

\end{itemize}


\section{Block~IX: Primordial-Black-Hole Domination\label{sec:block9}}


\begin{center}
{\bf Discussion Leader}\/: \\ Barmak Shams Es Haghi
\end{center}

PBHs are BHs which formed in the early universe in a fundamentally non-stellar way.  
They're interesting because they provide information about the early universe.  PBHs can 
also play the role of the DM, and they're a particularly interesting DM candidate because they 
require no new particle species beyond those of the SM.  They can be produced by the collapse 
of primordial overdensities in the early universe, by the dynamics of topological defects, 
by a scalar fifth force, by particle trapping by bubble walls, and by 
scalar-field fragmentation.  A PBH is characterized by its mass, spin, and charge.  
It is also interesting that if there is a non-standard cosmological history prior to BBN, the 
growth of PBHs can be enhanced without the need for fine-tuned features in the primordial 
power spectrum~\cite{Khlopov:2008qy,Clark:2016nst,Georg:2016yxa,Georg:2017mqk,Georg:2019jld}.

\begin{itemize}

\item Question: How generic {\it vs.}~fine-tuned is the production of PBHs?  

\item Answer: It depends a bit on the production mechanism. In the case of PBH formation 
due to the collapse of primordial overdensities, an enhancement in the power spectrum at 
small scales is necessary.  This, in turn, requires precise fine-tuning of the 
inflaton potential.  To avoid this fine-tuning, alternative mechanisms have been proposed, 
such as the confinement of a strongly-coupled sector involving heavy quarks~\cite{Dvali:2021byy}.

\end{itemize}

PBHs evaporate via Hawking radiation.  In general, a BH of mass $M_{\rm BH}$ acts like 
a thermal object (up to graybody factors) with a temperature 
\begin{equation}
  T_{\rm BH} ~=~ \frac{M_P^2}{8\pi M_{\rm BH}} ~\sim~ 
    10^{13} \left(\frac{1~{\rm g}}{M_{\rm BH}}\right)~{\rm GeV}~.
\end{equation} 
The mass range of interest for PBHs produced from the gravitational collapse of primordial
density perturbations after inflation is $0.1~{\rm g} < M_{\rm BH} < 10^9~{\rm g}$.  The lower 
limit of this range comes from the CMB constraint on the horizon size at the end of inflation. 
The upper limit ensures that the PBHs evaporate before BBN begins.  In what follows, I'll define 
$\beta$ to represent the initial abundance of PBHs.  There's a critical value $\beta_{\rm crit}$
of $\beta$ above which the PBHs will come to dominate the energy density of the universe before they
evaporate:
\begin{equation}
    \beta_{\rm crit} ~\sim~ \frac{M_P}{M_{\rm BH}}~.
\end{equation}

\begin{itemize}

\item Question: Is there a consensus on whether PBHs evaporate completely, 
or whether quantum-gravity effects come in and modify the dynamics of evaporation?

\item Answer: There is no agreement as to whether PBH evaporate completely or whether they 
leave a remnant.  

\item Comment: Yep.  I'll second that.  In order to address this point, we'd 
need theoretical tools for studying small black holes for which string and quantum-gravity effects 
need to be taken into account.

\end{itemize}

If the PBHs themselves do not constitute the DM, but if the DM interacts with the particles
of the visible sector only gravitationally, DM particles can nevertheless be produced as 
Hawking radiation by a population of evaporating PBHs.  Let's assume for concreteness that the 
mass spectrum of PBHs is monochromatic, or at least very sharply peaked around a particular mass 
$M_{\rm BH}$.  For such a PBH spectrum, one finds that the Hawking evaporation of PBHs with a 
small $\beta$ can account for the observed relic abundance of DM across a broad range of PBH 
and DM masses.  Interestingly, when $M_{\rm BH}$ is large and the mass $m_\chi$ of the DM is 
either very light ($m_\chi \sim 1$~MeV) or very heavy ($m_\chi\sim 10^9$~GeV or higher), the 
PBHs have to come to dominate the energy density of the universe in order to explain 
the observed relic abundance of DM.

\begin{itemize}

\item Question: How likely is it that the mass spectrum of PBHs is actually 
approximately monochromatic?  

\item Answer: If the PBHs form during a RD epoch, a monochromatic spectrum is not a 
unreasonable assumption.  By contrast, if they form during a MD epoch, matter-density 
perturbations grow once they enter the horizon, so one often gets a broader spectrum of 
masses, but this is model-dependent (a recent study in which this possibility is
examined extensively is Ref.~\cite{Gehrman:2023qjn}).  

\item Question: If there's a whole tower of DM 
particles~\cite{Dienes:2011ja,Dienes:2011sa}, can one produce the whole tower of them 
through evaporation?  One of the fundamental facets of the DM problem is the question
of how the DM is produced.  If one can produce many different kinds of DM particle democratically 
via a particular mechanism --- and it seems that, to a large extent, PBH evaporation is such a
mechanism --- one can more easily justify the production of multi-component DM.  

\item Answer: In principle, yes.  Of course, there's a lot one needs to account for in one's model 
of DM in order to determine which particles are going to be produced as Hawking radiation and in 
what relative amount.  This is basically determined by the particle masses relative to the initial 
temperature of the PBH in question.

\end{itemize}

The production of DM via PBH evaporation can affect the evolution of the DM abundance in other ways 
as well~\cite{Gondolo:2020uqv}.  If the DM also receives a contribution from thermal freeze-out, 
for example, the production of DM via PBH evaporation during the freeze-out epoch alters
the freeze-out dynamics by serving as a source of DM number density.  One can also have thermal 
and non-thermal DM production each yield separate contributions to the overall abundance --- for
example, if DM freezes out before the vast majority of the PBHs evaporate.  Thus, the interplay 
between PBH dynamics and the dynamics associated with various DM-production mechanisms can affect the
prospects for DM detection~\cite{Gondolo:2020uqv}.

\begin{itemize}

\item Question: How is the phase-space distribution of DM affected by the 
contribution from PBH evaporation in this case?  Do you get multiple peaks in $f(p)$ for the DM 
particles if the DM freezes out before the population of DM particles produced by PBH evaporation 
is generated and is population of PBHs never thermalizes?  Can the PBH population be warmer than 
the freeze-out population if it's produced later, after freeze-out?  Can the opposite be the case 
as well?  

\item Answer: It depends on the parameters.  Non-gravitational interactions allow DM particles 
to exchange energy with the thermal bath and to cool off.  As a result, you typically don't end 
up with the evaporation contributions to the DM phase-space distribution being sufficiently ``hot'' 
that they cause problems with large-scale structure due to free-streaming though.

\end{itemize}

Another interesting possible ramification of DM production from PBH evaporation is
that since this production mechanism is purely gravitational, it can --- and will --- also  
generically produce DM species that reside in a hidden sector.  In a lot of hidden-sector
DM scenarios, the visible and hidden sectors are both populated by inflaton decay.  It's 
often advantageous in such scenarios, due to constraints on dark radiation and such, to have 
the temperature in the hidden sector at the end of reheating be a lot lower than that of the 
visible sector, and there are ways of doing this~\cite{Hodges:1993yb,Berezhiani:1995am}.  
However, if the inflaton coulples to both the visible and hidden sectors, then inflaton-mediated 
processes will generically act to thermalize the visible and hidden sectors at later 
times~\cite{Adshead:2016xxj,Hardy:2017wkr,Adshead:2019uwj}.  Alternatively, one can imagine a 
different kind of scenario --- one in which only visible-sector particles are produced during 
reheating after inflation, but in which a cosmological population of hidden-sector particles is 
nevertheless generated at late times via PBH evaporation~\cite{Sandick:2021gew}.  If the 
dark/hidden-sector particles have non-negligible self-interactions, they can attain kinetic and 
chemical equilibrium and even potentially go through a cannibal phase.  When this is the 
case, the lightest hidden-sector particle can then potentially be produced by thermal freeze-out,
and if this particle is stable, it can play the role of the DM.  There's a range of $\beta$ for 
which achieving such a dark sector is possible.  In general, when you analyze the applicable 
constraints, you pretty much need $\beta < \beta_{\rm crit}$ in order to satisfy observational 
constraints, including Bullet-Cluster constraints~\cite{Randall:2008ppe}, on self-interactions.  

\begin{itemize}

\item Question: Are bounds from merging clusters like the Bullet Cluster really the 
leading bounds on these sorts of interactions?  What about the halo profiles of dwarf spheroidals 
and such?  

\item Answer: The constraints from all of these bounds are all roughly on the same order, 
so when one is performing a broad-brush analysis, one such bound is as good as another.  
It's therefore fine just to consider the Bullet-Cluster bound.  

\item Question: Can the hidden sector really be hotter than the visible 
sector?  It seems like the number of degrees of freedom --- \ie, $g_\ast(T)$ --- won't be 
enough to compensate.  

\item Answer: In these scenarios, a dark sector hotter than the visible sector is acceptable 
since it always happens when DM particles are non-relativistic and the dark sector can therefore 
be considered a ``cold'' sector.  Since the energy density associated with the particles in that 
sector behaves as massive matter, there isn't an impact on $H$.

\end{itemize}

PBH evaporation can also produce dark radiation~\cite{Arbey:2021ysg}.  When PBHs dominate the 
energy density of the universe (\ie, when $\beta > \beta_{\rm crit}$) and there are very light 
fields in the spectrum of your theory, there are constraints you need to contend with.

\begin{itemize}

\item Question: If you have huge numbers of light axion species, for example, 
and they're all being produced by PBH evaporation, it would seem like this would cause 
severe phenomenological problems.  Is this indeed the case?

\item Answer: Indeed, this is the case.  You can place bounds on the number of light axion 
species in scenarios with a PBH-dominated epoch.  In particular, there are 
$\Delta N_{\rm eff}$-related constraints on dark radiation from PBH evaporation, and these
constraints yield a bound $N_{\rm ax} \lesssim 7$ on the number of light axion species that
can be present in the theory if PBHs dominate the universe~\cite{Hooper:2019gtx}.  In other words, 
if such a PBH epoch is present in the cosmological timeline, there's no axiverse, no stasis epoch 
involving these light particles subsequent to the epoch of PBH-domination, \etc~  

\item Question: If you have a light boson, won't the production be dominated 
when its Compton wavelength is comparable to the PBH radius?   

\item Answer: Yes, that's the regime in which you get superradiance.  Within this regime, then, 
there's a contribution to the production rate of such light particles from superradiance and 
another contribution from Hawking evaporation.  There are two regimes within which superradiance 
can happen: one in which the BHs are very heavy and the bosons are very light and one in which the 
PBHs are very light and bosons are very heavy. 

\item Question: Can you calculate a velocity dispersion for the PBHs produced by 
mergers, or are the random velocities so small that their velocity dispersion doesn't matter?  

\item Comment: The peculiar velocities are typically small and can be 
ignored, especially since you're averaging over random velocities of the stuff that collapses 
to form a PBH when it forms from gravitational collapse.  In practice, people usually just 
look at the spatial distribution of PBHs and examine how likely two nearby PBHs are to form a 
gravitationally bound system, and then examine how probable it is that this system is perturbed 
by a third object.

\item Question: How long does it take to generate a sizeable population 
of spinning PBHs via mergers?  

\item Answer: People have looked at this~\cite{Fishbach:2017dwv}.  In general, 
in the situations I've talking about here, a sizeable population can be generated within a 
cosmologically reasonable time span.  To ensure that mergers occur successfully before the 
evaporation of PBHs, the PBH capture and merger rates must be comparable to the expansion 
rate~\cite{Zagorac:2019ekv}.

\item Question: Can you generate a cosmological population of relativistic PBHs 
somehow?

\item Answer: It's an interesting idea, but I haven't found a way of doing it.  

\item Comment: When soliton-like configurations merge, unlike PHBs, 
there can be partial mergers which create a smaller, rapidly moving remnant in addition to 
the larger merged object.  These remnants aren't relativistic, but they can have large velocities 
in comparison with the rest of the solitons in the soliton ``gas.''

\item Comment: You can also get velocity ``kicks'' due to many-body BH interactions, 
such as when a BH binary interacts with a third BH~\cite{Delos:2024poq}.  Essentially, the orbital 
motion of the binary can be converted into the streaming motion of a free BH.  However, BHs can't 
really become ultrarelativistic as a result of these kicks, since the binary orbital motion is not 
itself ultrarelativistic.

\item Question: What about bubble-wall collisions?  Can a PBH inherit a net 
momentum from the colliding bubbles?  

\item Answer: Yes, but again, getting a relativistic velocity for a PBH in this manner is 
probably challenging.

\end{itemize}

There are many ways in which PBHs can give rise to GW signals.  One is that GWs can be 
produced in association with processes that produce PBHs.  For example, scalar-induced GWs 
from primordial 
overdensities, or GWs can be produced during first-order phase transitions which are likewise 
responsible for PBH formation.  Another way in which PBHs can give rise to GWs is via 
the mergers then can undergo prior to their evaporation~\cite{Zagorac:2019ekv}.  Yet third way 
in which PBHs can give rise to GWs is via Hawking radiation
directly~\cite{Anantua:2008am,Dolgov:2011cq,Dong:2015yjs,Ireland:2023avg}.  A fourth and final
way in which PBHs can give rise to GWs is via the rapid oscillations of sub-horizon modes while
the universe is undergoing a ``rapid'' transition from a period of PBH-domination (an EMDE) to 
a RD epoch.  As others have already commented on at several points during this workshop,  
these oscillations can lead to an enhancement in GW emission~\cite{Inomata:2019ivs,Pearce:2023kxp}.  
For very high-frequency GWs, one promising detection method is to take advantage of the inverse 
Gertsenshtein effect~\cite{Gertsenshtein:1962xxx}, which involves the conversion of GWs into 
photons in the presence of a $B$-field.  This detection method is particularly promising for 
GWs with frequencies in the MHz -- GHz range~\cite{Li:2009zzy,Ringwald:2020ist}.  GWs within this
frequency range can be generated through a number of means, including early phase 
transitions in the very early universe, oscillon and cosmic-string dynamics, and preheating.
There can also can be an intimate connection to baryogenesis in scenarios in which the 
baryon asymmetry is generated by the decays of heavy particles produced via PBH 
evaporation~\cite{Gehrman:2022imk}.  A similar correlation exists between  MHz -- GHz GWs 
and DM production by PBH evaporation~\cite{Gehrman:2023esa}.

\begin{itemize}

\item Question: Since it's relevant here, what's the status 
of CMB-S4 at this point?  

\item Answer: The party line seems to be that construction of the telescope 
in Chile will still move forward, but that the timeline will be delayed by a couple of years.  
The telescope in Antarctica will in principle eventually still be built at some point, but
construction will be delayed by a decade, but who knows?  

\item Comment: If the telescope at the South Pole doesn't move forward, a reduction 
in our ability to measure the tensor-to-scalar ratio accurately will be the most significant 
ramification.  

\item Comment: One contribution that's being indicated on the plots that have
been shown during this block is the Poltergeist effect~\cite{Inomata:2019ivs}, which is a 
``spiking up'' a pre-existing GW spectrum in the presence of PBHs.

\item Question: There's a lot of interest in BHs from a formal-theory perspective --- 
entanglement entropy, Page curves, and so forth --- and a lot of phenomenological interest in BHs 
as well.  However, the formal-theory and phenomenology communities don't talk to each other much.  
Are there interesting conversations/interfaces between the two communities to be had with regard to
this topic?  Is there anything formal theorists can do that might shed light on phenomenological BH
calculations?  

\item Answer: There are lots of things to be done at the potentially fruitful 
interface between formal theory and BH phenomenology.  In practice, however, formal-theory 
calculations are best performed in controlled settings --- \eg, with BHs which are extremal 
(or near extremal), in theories with supersymmetry, in low-dimensional gravity theories, or in 
gravitational systems with known holographic duals, \etc~  Fewer concrete results have been derived 
in more realistic settings where some of these theoretical crutches are absent.  For instance, 
the Schwinger production rate for an extremal Reissner-Nordstr\"{o}m black hole in the 
full geometry has only recently been computed~\cite{Lin:2024jug} without recourse to supersymmetry 
or holography.  Nonetheless, it is conceivable that some of these theoretical insights continue to 
apply in less controlled settings.

\end{itemize}


\section{Summary Discussion and Workshop Synthesis\label{sec:summary}}


\begin{center}
{\bf Discussion Leaders}\/: \\ Brian Batell, Keith R.~Dienes, Brooks Thomas, and Scott Watson
\end{center}

This has been a great tour through many of the ideas surrounding non-standard 
cosmological epochs and expansion histories.   During this final session, we thought 
that it would be appropriate to pose a number of questions for all of us to ponder, 
questions which reach across all of the different scenarios we have discussed 
individually, and assess the comparative properties and potential abilities of 
these different scenarios to address the overall themes of this workshop. 
\bigskip

Our first question is to ask what you all think are  the most promising ways we have for 
distinguishing between different cosmological expansion histories prior to BBN?

\begin{itemize}
  \item  Answer: The matter power spectrum, provided the DM 
    is not completely cold.
  \item Answer: The cosmic neutrino background is also promising if
    it can be detected and characterized.
\end{itemize}

Okay, that makes sense.  So maybe we can now consider what might be the most 
constructive ways in which formal theory and the top-down perspective it brings 
can inform the effort to understand the expansion history of our universe?

\begin{itemize}
  \item Answer: Predictions for the GW background from specific top-down 
      models would be valuable.
  \item Answer: I respect string theory very deeply, but we need to acknowledge 
    that there is a vacuum-selection problem.  Progress on issues related to this problem would 
    help to indicate which parts of the landscape are more probable than others.
  \item Answer: The swampland program is a step in that direction. 
  \item Comment: If something like the Hubble tension survives, what does it 
    tell us about top-down theories?  That's a question that probably deserves more 
    attention.
\end{itemize}

Are there any observational ``smoking guns'' that would point the way to
particular modifications of the expansion history?

\begin{itemize}
  \item Answer: It's hard to know what a ``smoking gun'' is.  One example of 
    a smoking gun for non-standard expansion histories in general might be the Hubble tension.  
    However, there are many ways to address it.
  \item Comment: I disagree that the Hubble tension was ever a smoking gun.  The 
    values from early- and late-universe measurements should have been significantly different.  
    They weren't.
  \item Comment: We should vote on whether the Hubble tension is real.
  \item Comment: Systematics are typically not taken into account when people 
    quote the discrepancy between early- and late-universe measurements.
  \item Comment: I disagree with that statement.
  \item Comment: We shouldn't vote.  I'm not an observer and I don't feel qualified 
    to provide a critique.  Also, on a different topic, non-linear regimes need 
    to be studied and better understood.
  \item Question: Someone raised an interesting question a moment ago.  More generally, 
    should we expect departures from $\Lambda$CDM to be large or small?
  \item Answer: Cosmology pre-BBN is the wild west, whereas post-BBN cosmology is 
    well understood. 
  \item Comment: I actually agree with the earlier comment that the Hubble 
    tension could be a smoking gun signal that our universe had a non-standard expansion history. 
    There's certainly wiggle room to modify the expansion history after BBN.  However, the 
    modification needs to be small, so it would show up first as a tension between low-redshift 
    observations and the expectations set by the CMB. 
\end{itemize}

To follow up on the previous question, what sorts of features in the GW background might we 
feel more confident about identifying as signals of non-standard cosmology --- as 
opposed to astrophysics (\eg, BHs)?
\begin{itemize}
  \item Answer: GWs offer one of the best opportunities to characterize the 
    cosmic-expansion history.  In that respect, GW detectors are analogous to high-energy 
    colliders.
  \item  Comment: There's wiggle room after BBN for 
    minor modifications of the expansion history, but the effect on the GW background would be small. 
  \item  Question: From what source in observational cosmology do we think 
    we have the best measurements?  We might expect deviations to show up there.  Perhaps we
    might hope to see them in CMB data, since not only do we measure the CMB incredibly precisely, 
    but the physics which affects the CMB is also linear, so we can be confident in our predictions.
  \item Answer: Indeed.  There will be new CMB data forthcoming from 
    ACT~\cite{ACT:2020frw}, \etc, which will improve our understanding of the Hubble tension and 
    other related tensions (\ie, the cosmic calibration tension).
\end{itemize}
 
To what extent can the expansion history be modified after BBN? How compelling is the 
need to do this in light of the (to use the helpful phrase Tristan [Smith] introduced earlier 
in this workshop) cosmic-calibration tension?
\begin{itemize}
  \item Question: Wouldn't it be shocking that there were late time 
    modifications to $\Lambda$CDM?
  \item Answer: Maybe not.  What about $\Delta N_{\rm eff}$, for example?
  \item Comment: New physics for addressing the Hubble tension and other related 
    tensions is very peculiar. Why such a small effect?
\end{itemize}

To what extent are non-perturbative effects unavoidable during reheating?  Are there 
any general --- even if qualitative --- lessons to be learned about the way inflation ends 
and the universe becomes radiation-dominated again --- other than perhaps that such 
non-perturbative effects generally play a substantial role in the process?
\begin{itemize}
  \item Answer: I think there are some general statements that can be made.  Homogeneous, oscillating 
  scalar fields are unstable to spatial fragmentation, either due to self-interactions or due to 
  gravitational interactions.  So, even in single-field inflation models with very small coupling to 
  other fields, the end state of the field after a sufficiently long time interval will be inhomogeneous on 
  subhorizon scales.  When it couples to other bosonic fields, the rolling or the coherent oscillations 
  of the scalar field will generally lead to more efficient energy transfer due to non-perturbative effects 
  in comparison with the perturbative case.  However, back-reaction from the fields produced in this
  way can limit efficient production in certain cases.  Finally, radiation domination can be achieved 
  without fields being in a thermal state, so care should be taken in distinguishing between different
  kinds of RD epochs. 
\end{itemize}

What are the most constructive ways in which terrestrial experiments can inform our 
understanding of what the universe was like prior to the BBN epoch?  More broadly speaking, 
what are the most exciting experimental/observational probes of the expansion history?
\begin{itemize}
  \item Answer: We have particle-physics theories that are useful for 
    understanding BBN.  In the future, we'd presumably like to move to developing theories 
    that are useful for understanding what happened at higher energies and earlier times.
  \item Comment: DM physics could give us a probe into physics at earlier times --- 
    for example, if an axion is discovered by means of an axion-photon coupling.
  \item Comment: GWs are extremely exciting.  Everything couples to gravity, 
    so there can be loads of new hidden sectors (perhaps those predicted by string theory) 
    which couple to gravity but don't couple directly to the visible sector.  Gravitational waves 
    are the way to detect such physics, since gravity provides the only other ``spectrum'' 
    (beyond the electromagnetic spectrum) through which we can look at the universe. 
  \item Comment: Ultra-high-energy cosmic rays provide another promising probe of 
    early-universe cosmology, and potentially of very-high-energy physics.  
\end{itemize}

Okay, thanks!   This has been a very interesting discussion.

We have now reached the end of the workshop.   People who wish to linger and continue the conversation 
are welcome to do so.   We still have plenty of food left, so please help yourselves and have a safe 
trip back home.  



\begin{acknowledgments}

This workshop was attended by many more people than just the organizers and 
the Discussion Leaders.  In this context, we are particularly 
happy to acknowledge the insightful comments of the other registered workshop participants, 
specifically Ali Beheshti, Amit Bhoonah, Francis Burk, Shuyang Cao, Linda Carpenter, 
Morgan Cassidy, Kun Cheng, Chris Choi, Swapnil Dutta, Brenda Gomez Cortes, Murman Gurgenidze, 
Marcell Howard, Wenjie Huang, Tina Kahniashvili, Juhun Kwak, Monica Leys, Matthew Low, 
Jacob Magallanes, Sayan Mandal, Elizabeth Meador, Jeffrey Newman, Sneh Pandya, 
Arnab Pradhan, Zahra Tabrizi, Si Wang, Arthur Wu, Jaeok Yi, and Andrew Zentner.
The authors of this document wish to thank everyone for 
their lively comments and questions throughout the workshop, all of which added to 
the richness and dynamic nature of the proceedings.  The authors of this document would also like to acknowledge the Pittsburgh Particle Physics 
Astrophysics and Cosmology Center (PITT-PACC) for providing both financial and 
organizational support for this workshop.  We would also like to thank the other members of the 
local organizing committee (specifically Amit Bhoonah, Kun Cheng, Matthew Low, and Zahra Tabrizi) 
and the support staff (Chris Condon, Joni George, and Gracie Gollinger) at the University of 
Pittsburgh for their critical assistance during the workshop.  Finally, we would like to thank 
Tao Han, whose leadership, wisdom, advice, and support were instrumental in making this 
non-traditional workshop a reality.

The research activities of RA are supported in part by the U.S.\ National 
Science Foundation under Grant PHY-2210367.
The research activities of MA are supported in part by the U.S.\ Department of 
Energy under Grant DE-SC0021619.
The research activities of BB are supported in part by the U.S.\ Department of 
Energy under Grant DE–SC0007914.
The research activities of KRD are supported in part by the U.S.\ National 
Science Foundation through its employee IR/D program as well as by the U.S.\ 
Department of Energy under Grant DE-FG02-13ER41976 / DE-SC0009913.
The research activities of ALE are supported in part by the U.S.\ National 
Science Foundation under Grant PHY-2310719.
The research activities of AG are supported in part by U.S.\ Department of Energy 
under Grant DE–SC0007914 and the GRASP initiative at Harvard University.
The research activities of JTG are supported in part by the U.S.\ National Science 
Foundation under Grant PHY-2309919.
The research activities of JH are supported in part by the U.S.\ National 
Science Foundation under NSF CAREER Grant PHY-1848089 and under NSF Cooperative Agreement 
PHY-2019786 (The NSF AI Institute for Artificial Intelligence and Fundamental Interactions).
The research activities of FH are supported in part by ISF Grant 1784/20 and by 
MINERVA Grant 714123.
The research activities of AJL are supported in part by the U.S.\ National 
Science Foundation under Grant PHY-2412797.
The research activities of BSE are supported in part by the U.S.\ Department of
Energy under Grant DE-SC0022021.
The research activities of JS are supported in part by the U.S.\ Department of
Energy under Grant DE-SC0015655.
The research activities of GS are supported in part by the U.S.\
Department of Energy under Grant DE-SC0017647..
The research activities of KS are supported in part by the U.S.\ National 
Science Foundation under Grant PHY-2412671.
The research activities of TLS are supported in part by the U.S.\ National
Science Foundation under Grants AST-2009377 and AST-2308173.
The research activities of BT are supported in part by the  U.S.\ National 
Science Foundation under Grant PHY-2014104.  BT would like to thank Tao Han 
and the University of Pittsburgh for hospitality.
The research activities of SW are supported in part by the U.S.\ Department of
Energy under Grant DE-FG02-85ER40237.  SW would like to thank Alexey Petrov and 
the University of South Carolina for hospitality.
The opinions and conclusions expressed herein are those of the authors, and do not 
represent any funding agencies. 

\end{acknowledgments}

\bibliography{references}
\end{document}